\documentclass[usenatbib,preprint]{mn2e}

\usepackage[switch]{lineno}
\usepackage{graphicx}
\usepackage {natbib,aas_macros}
\usepackage {mediabb}
\usepackage {bm}
\usepackage {amsmath,amssymb}

\newcommand{\msun}{\thinspace M_\odot} 
\newcommand{\gcm}{~{\rm g~cm}^{-3} }

\newcommand{\magB}{\mathbf{B}}

\newcommand{\cul}{\mathbf{J}}
\newcommand{\vel}{\mathbf{v}}
\newcommand{\rad}{\mathbf{r}}
\newcommand{\etacm}{~{\rm cm}^{2} ~{\rm s}^{-1} }

\newcommand{\ms}{~{\rm m} ~{\rm s}^{-1} }

\newcommand{\mum}{{\rm \mu} {\rm m} }
\newcommand{\mm}{{\rm mm}}
\newcommand{\nm}{{\rm nm}}
\newcommand{\cm}{{\rm cm}}
\newcommand{\au}{{\rm AU}}
\newcommand{\g}{{\rm ~g}}

\newcommand{\msunyear}{\thinspace M_\odot~{\rm yr}^{-1}}

\newcommand{\dv}{\Delta \vel}

\newcommand{\tstop}{t_{\rm stop}}

\title{Co-evolution of dust grains and protoplanetary disks}

\author[Tsukamoto et al]{
Yusuke Tsukamoto$^{1}$, Masahiro N. Machida$^{2}$, and  Shu-ichiro Inutsuka$^{3}$ \\
$^1$Graduate Schools of Science and Engineering, Kagoshima University, Kagoshima, Japan  \\
$^2$Department of Earth and Planetary Sciences, Kyushu University, Fukuoka, Japan \\
$^3$Department of Physics, Nagoya University, Aichi, Japan  \\
}

\begin{document}
\maketitle

\begin{abstract}
We propose a new evolutionary process of protoplanetary disks "co-evolution of dust grains and protoplanetary disks", revealed by dust-gas two-fluid non-ideal magnetohydrodynamics simulations considering the growth of dust and associated changes in magnetic resistivity. We found that the dust growth significantly affects disk evolution by changing the coupling between the gas and magnetic field. Moreover, once the dust grains sufficiently grow  and the adsorption of charged particles on dust grains becomes negligible, the physical quantities (e.g., density and magnetic field) of the disk are well described by characteristic power laws. In this disk structure, the radial profile of density is steeper and the disk mass is smaller than those of the model ignoring dust growth.
We analytically derive these power laws from the basic equations of non-ideal magnetohydrodynamics. The analytical power laws are determined only by observable physical quantities, e.g., central stellar mass and mass accretion rate, and do not include difficult-to-determine parameters e.g., viscous parameter $\alpha$.  Therefore, our model is observationally testable and this disk structure is expected to provide a new perspective for future studies on protostar and disk evolution.

\end{abstract}

\begin{keywords}
star formation -- circum-stellar disk -- methods: magnetohydrodynamics -- smoothed particle hydrodynamics -- protoplanetary disk
\end{keywords}

\section{Introduction}
\label{intro}

In protoplanetary disks, dust grains are not only the building blocks of the planets, but also play a key role in determining the ionization degree of the disk gas by adsorbing charged particles (ions and electrons) in the gas phase. Since the ionization degree determines the magnetic resistivity, i.e., the degree of coupling between the gas and magnetic field, the microscopic nature ($\mum$ to $\cm$ scale) of the dust grains is expected to influence the macroscopic disk evolution ($100$ AU, i.e., $10^{15} \cm$ scale) via magnetic resistivity \citep{2016MNRAS.460.2050Z,2020ApJ...900..180M,2020A&A...643A..17G,2022ApJ...934...88T}.

Previous studies show that non-ideal magnetohydrodynamics (MHD) effects arising from finite resistivity, specifically ambipolar diffusion, dramatically weaken the coupling between the magnetic field and gas, thereby enabling the formation of a disk \citep{2015ApJ...801..117T, 2015MNRAS.452..278T, 2016MNRAS.457.1037W,2016A&A...587A..32M}.
Moreover ambipolar diffusion determines the magnetic flux evolution in protostars \citep{1998ApJ...497..850L,2020ApJ...896..158T}.

Previous studies on the formation and evolution of protoplanetary disks assumed that dust grains possess the properties (such as the size distribution) of  the interstellar medium (ISM) dust grains. However, in the disk, the dust growth timescale is $\sim 10^4$ years, which is much shorter than the lifetime of the disks ($\sim 10^6$ years). Thus, it is unsatisfactory to study disk evolution with resistivity assuming ISM dust \citep{2022arXiv220913765T}.

How would the dust growth affect the ionization degree? As dust grains merge and grow, their total surface area decreases.
Therefore, the adsorption of charged particles by the dust grains becomes ineffective, and the gas-phase ionization degree is expected to increase and magnetic resistivity to decrease accordingly. Recent studies on dust growth and associated changes in magnetic resistivity have shown a decrease in magnetic resistivity \citep{2016MNRAS.460.2050Z,2020ApJ...900..180M,2020A&A...643A..17G,2022ApJ...934...88T, 2022MNRAS.515.2072K}, and some studies have also shown changes in gas dynamics as a result \citep{2023MNRAS.518.3326L,2023arXiv230101510M}.

However, the effect of the dust growth on the evolution of the protoplanetary disk is still unclear, because the calculations in the aforementioned studies were performed assuming spherical symmetry \citep{2023MNRAS.518.3326L} or 3D simulation until the prestellar or first core formation stage in which the gas is supported by the pressure gradient force \citep{2023arXiv230101510M}.

In this study, we report simulation results of the formation and evolution of protoplanetary disks of $\sim 10^4$ years after the formation of protostars considering dust growth inside the disk, the associated change of magnetic resistivity, and its feedback on the disk dynamics.
Moreover we present an analytical argument that explains the resulting disk structures.
Based on these results, we propose a new evolutionary process for protostars: "co-evolution of dust grains and protoplanetary disks".

\section{Methods and initial condition}
\subsection{Numerical methods}
\subsubsection{Two-fluid magneto-hydrodynamics simulations}
We solve two-fluid magnetohydrodynamics equations for the dust-gas mixture.
The governing equations are given as
\begin{eqnarray}
  \label{single_fluid_continum}
  \frac{D \rho}{D t}  &=& - \rho \nabla \cdot \vel, \\
  \label{single_fluid_eps}
  \frac{D \epsilon}{D t} &=& -\frac{1}{\rho} \nabla \cdot \{\epsilon(1-\epsilon)\rho \Delta \vel \}, \\
  \label{single_fluid_motion1}
  \frac{D \vel }{D t} &=&  -\frac{1}{\rho} \{\nabla P-  \frac{\cul \times \magB}{c} \}, \\ 
  \label{single_fluid_dv2}
  \frac{D \dv }{D t} &=&-\frac{\dv}{\tstop}+\frac{1}{\rho_g}[-\nabla P+ \frac{\cul \times \magB}{c}],\\
  \frac{D \magB}{Dt}&=&-\magB (\nabla \cdot \vel)+(\magB\cdot \nabla)\vel \nonumber \\
  &+& c \nabla \times \{  \eta_{O} \cul +\eta_A (\cul \times {\hat \magB}) \times {\hat \magB}  \},
\end{eqnarray}
where $\rho_{[g, d]}$ denotes the mass densities and subscripts $[g, d]$ denote gas and dust components, respectively. $\rho=\rho_g+\rho_d$ denotes the total density.
$\epsilon=\rho_d/\rho$ denotes the dust-to-total-mass ratio,
$\vel=(\rho_g \vel_g+\rho_d \vel_d)/(\rho_g+\rho_d)$  denotes the barycentric velocity of dust gas mixture where $\vel_{[g, d]}$ denotes the gas and dust velocity,
$\dv=(\vel_d-\vel_g)$  denotes the the velocity difference between gas and dust,
$P$  denotes the gas pressure.
$\cul$  denotes the electric current.
$c$  denotes the speed of light.
$\eta_O$ and $\eta_A$   denote the Ohmic and ambipolar resistivities, respectively.
The details of the approximations adopted in the governing equations and numerical method are described in \citet{2021ApJ...913..148T}.
Our numerical simulations consider the Ohmic and ambipolar diffusions, but ignore the  Hall effect.

We adopted a barotropic equation of state (EOS)
in which the gas pressure  depends only on the density.
\begin{eqnarray}
  \label{eos_eq}
  P &=& P(\rho_g)=
  \begin{cases}
    c_{s, ref}^2 \rho_g \left\{ 1+\left(\frac{\rho_g}{\rho_c}\right)^{\frac{2}{3}} \right\}& (\rho_g<\rho_e)\\
        c_{s, ref}^2 \rho_g \left\{ 1+\left(\frac{\rho_e}{\rho_c}\right)^{\frac{2}{3}}\left(\frac{\rho_g}{\rho_e}\right)^{\frac{2}{5}}\right\} & (\rho_g\ge \rho_e)\\
      \end{cases}
\end{eqnarray}
$c_{s, ref}=190 \ms$ is isothermal sound velocity at $10$ K.
We used a critical density of $\rho_c=4\times 10^{-14} \gcm$ above which gas behaves adiabatically and $\rho_e= 10^{-11} \gcm$ above which gas behaves as diatomic molecule.
In our simulations, the gas density is in the range of $\rho_g<10^{-9}\gcm$ and ignoring the dissociation of H$_2$ in the equation of state does not affect the results.

\subsubsection{Dust growth}
We consider dust growth
with single-size approximation \citep{2016A&A...589A..15S,2016ApJ...821...82O, 2017ApJ...838..151T}.
The governing equation of dust growth is
\begin{eqnarray}
  \label{dadt}
  \frac{D_d a_{d}}{Dt}=A_{\rm gain/loss} \frac{a_d}{3 t_{\rm growth}},
\end{eqnarray}
where $a_{d}$ denotes the representative dust size, $t_{\rm growth}=1/(\pi a_d^2 n_d \Delta v_{\rm dust})$, $n_d$ denotes the dust number density,
$\Delta v_{\rm dust}$ denotes the collision velocity between the dust grains,
$D_d/Dt=\partial/\partial t +\vel_d \cdot \nabla$,
and
\begin{eqnarray}
  \label{collision_factor}
A_{\rm gain/loss}={\rm min}(1,-\frac{\ln(\Delta v_{\rm dust}/\Delta v_{\rm frag})}{\ln 5}),
\end{eqnarray}
which models the collisional mass gain and loss \citep{2016ApJ...821...82O}.
We assume $v_{\rm frag}=30 \ms$.
For the dust relative velocity $\Delta v_{\rm dust}$, we consider
the sub-grid scale turbulence and Brownian motion.

For the turbulent-induced dust relative velocity $\Delta v_{\rm turb}$,
we adopt the prescription presented by \citet{2007A&A...466..413O},
\begin{align}
\Delta v_{\rm turb}&= \nonumber \\
&\begin{cases}
  \frac{\delta v_{\rm Kol}}{t_{\rm Kol}} (t_{\rm stop, 1}-t_{\rm stop, 2}) & (t_{\rm stop, 1}<t_{\rm Kol}) \nonumber \\
  1.5 \delta v_L \sqrt{\frac{t_{\rm stop, 1}}{t_L}} & (t_{\rm Kol}< t_{\rm stop, 1}<t_L)  \nonumber \\
  \delta v_L \sqrt{\frac{1}{1+t_{\rm stop, 1}/t_L}+\frac{1}{1+t_{\rm stop, 2}/t_L}} & (t_L< t_{\rm stop, 1})  \nonumber 
\end{cases}
\\
\end{align}
where  $\delta v_{\rm Kol}=Re_L^{-1/4}$ and $t_{\rm Kol}=Re_L^{-1/2} t_L$ denote the
eddy velocity and eddy turn-over timescale at dissipation scale and $Re_L=L v_L/\nu$ denotes the Reynolds number.
We set the stopping time $t_{\rm stop, 1}=t_{\rm stop}(a_d)$
and  $t_{\rm stop, 2}=1/2~ t_{\rm stop}(a_d)$ referring to \citet{2016A&A...589A..15S}, where
$t_{\rm stop}(a_d)$ is calculated as in our previous study \citep{2021ApJ...913..148T}.

We assume sub-grid turbulence of the ``$\alpha$ turbulence model "\citep{2021ApJ...920L..35T}, in which we assume
\begin{eqnarray}
  \delta v_L&=&\sqrt{\alpha_{\rm turb}} c_s, \\
  \label{tL_eq}
  t_L&=&\frac{c_s}{a_g}, \\
  L&=&\delta v_L t_L.
\end{eqnarray}
$\alpha_{\rm turb}=2 \times 10^{-3}$ denotes the dimensionless parameter that determines the strength
of the sub-grid turbulence and  $a_g$ denotes the gravitational acceleration.
(see \citet{2021ApJ...920L..35T} for the underlying physical assumptions for this turbulence model).
  $\alpha_{\rm turb}$ is difficult to be determined from simulations, while it does not affect the dust growth timescale as strongly as density.
 In this paper, we adopted this fixed value just for simplicity.  
 Larger (smaller) $\alpha_{\rm turb}$ value decreases (increases) the dust growth timescale.

\subsubsection{Resistivity calculations}
For the resistivity model, we adopt the analytical resistivity formula described in \citet{2022ApJ...934...88T} in which we analytically solve the  equations for chemical equilibrium in the gas phase and detailed balance equations for dust charging. 
The dust size distribution considered in resistivity calculations is set to be
\begin{eqnarray}
  \label{power_ad}
  \frac{{\rm d} n_{\rm d}}{{\rm d} a} =  A~ a^{-q}  (a_{\rm min}<a<a_{\rm max}),
\end{eqnarray}
where $A$ denotes a constant for normalization.
Here we assume that the maximum dust size is $a_{\rm max}=a_{\rm d}$ of equation (\ref{dadt}) (which is valid when $q<4$).
The minimum dust size $a_{\rm min}$ and power exponent $q$ are the parameters of this study.

  The temperature for the resistivity calculation is calculated according to the equation of state (equation (\ref{eos_eq})).

  Our resistivity model does not include charge neutralization due to grain-grain collisions because
  the dust grains tend to coalesce and grow rather than bounce when (sub-)micron-sized small dust particles collide \citep{1997ApJ...480..647D,2000PhRvL..85.2426B,2012Icar..218..688W,2015ApJ...798...34G}.
  Note also that the grain-grain neutralization is important for resistivities only when there are significant amount of small dust grains and they contribute to the electric current. Since we are interested in the impact of dust growth and in the situations that the contribution of the dust grains becomes minor, neglecting grain-grain neutralization does not change our main claims in this paper.
  See \citet{2022ApJ...934...88T} for more discussions.

\subsubsection{Sink particle}
A sink particle is dynamically introduced when the density exceeds $\rho_{\rm sink}=10^{-12} \gcm$.
The sink particle absorbs SPH particles with $\rho>\rho_{\rm sink}$ within $r_{\rm sink}<1$ AU.

\subsection{Initial conditions}
We adopt the density-enhanced Bonnor-Ebert sphere, which is surrounded by a medium with a steep density profile used in \citet{2021ApJ...920L..35T}  as a initial condition. 

The radius of the core
is $R_c=4.8\times 10^3$ AU and the enclosed mass within $R_c$ is $M_c=1 \msun$.
We adopt an angular velocity  profile of $\Omega(d)=\Omega_0/[\exp[10(d/(1.5 R_c))-1]+1]$
with $d=\sqrt{x^2+y^2}$ and  $\Omega_0=2.3\times 10^{-13} ~{\rm s^{-1}}$.
We assume a constant magnetic field $(B_x,B_y,B_z)=(0,0,50 ~\mu G)$.

The parameter $\alpha_{\rm therm}$ ($\equiv E_{\rm therm}/E_{\rm grav}$) is $0.4$, where
$E_{\rm therm}$ and $E_{\rm grav}$ denote the thermal and gravitational energies of the central core
(without surrounding medium), respectively.
The parameter $\beta_{\rm rot}$ ($\equiv E_{\rm rot}/E_{\rm grav}$) within the core is $0.03$,
where $E_{\rm rot}$ denotes the rotational energy of the core.
The mass-to-flux ratio of the core normalized by the critical value is  $\mu/\mu_{\rm crit}=5$.

We adopt a dust density profile of $\rho_d(r)=f_{dg} \rho_g(r)/[\exp[10(r/(1.5 R_c))-1]+1]$,
where $f_{dg}=10^{-2}$ denotes the dust-to-gas mass ratio.
The dust density profile  has the same shape
with the gas density profile in $r\lesssim 1.5 R_c$
but is truncated at $r \geq 1.5 R_c$. The initial (maximum) dust size is assumed to be $a_d=0.1 \mum$.

We resolve 1 $\msun$ with $3\times 10^6$ SPH particles.
The model names and parameters of the models are listed in Table 1.

\begin{table*}
\label{Model_table}
\begin{center}
  \caption{
    Model name, minimum dust size $a_{\rm min}$, power exponent for dust size distribution $q$, and cosmic ray ionization rate $\zeta_{\rm CR}$.
    "Y" means that the dust growth is considered in the simulation and "N" means that the dust growth is not considered.
}		
\begin{tabular}{ccccc}
\hline\hline
 Model name  & $a_{\rm min} [\nm]$ &$q$& $\zeta_{\rm CR} [{\rm s^{-1}}]$ & Dust growth  \\
\hline
ModelA100Q25 & 100  & 2.5 & $10^{-17}$  & Y \\
ModelA100Q35 & 100  & 3.5 & $10^{-17}$  & Y \\
ModelA5Q25 & 5      & 2.5 & $10^{-17}$  & Y \\
ModelA5Q35 & 5      & 3.5 & $10^{-17}$  & Y \\
ModelZeta18& 100    & 2.5 & $10^{-18}$  & Y \\
ModelA100Fixed & 100& -   & $10^{-17}$  & N \\
\hline
\end{tabular}
\end{center}
\footnotesize
\end{table*}

\section{Results}
\subsection{Co-evolution of dust grains and protoplanetary disks}

\begin{figure*}
  \includegraphics[clip,trim=0mm 0mm 0mm 0mm,width=150mm]{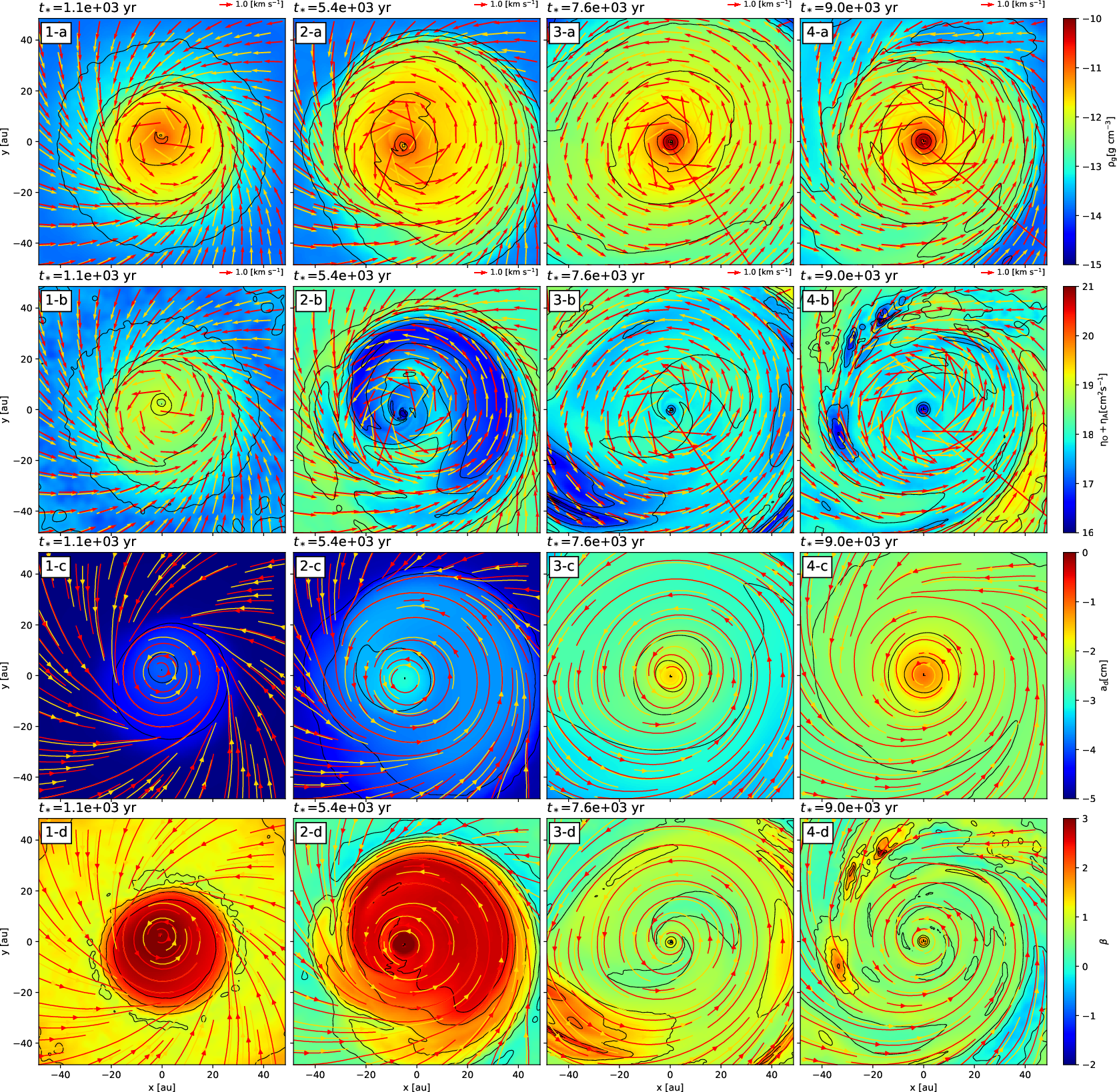}
  \caption{
    Time evolution of the (a) gas density, (b) magnetic resistivity ($\eta_O+\eta_A$), (c) dust size ($a_d$), and  (d) plasma $\beta$ on the $x$-$y$ plane with $100$ AU box.
    The time after protostar formation are shown in the upper left.
    The red and orange arrows show the velocity field of the dust and gas, respectively, on the $x$-$y$ plane.
    The red and orange lines show their respective streamlines.
    The black lines are the contour of the quantity of each panel.
    The contour levels are (a) $\rho_g=10^{-15},10^{-14.5},\cdots,10^{-10}\gcm$, (b) $\eta_O+\eta_A=10^{16},10^{16.5},\cdots,10^{21}\etacm$,
    (c) $a_d=10^{-5},10^{-4.5},\cdots,10^{0}\cm$, and (d) $\beta=10^{-2}, 10^{-1.5},\cdots, 10^{3}$.
  }
  \label{2D_time_evolution_q25_a100nm}
\end{figure*}

Figure \ref{2D_time_evolution_q25_a100nm} shows the time evolution of (a) density, (b) total magnetic resistivity ($\eta_O+\eta_A$), (c) dust size, and (d) plasma $\beta$ in our fiducial model, ModelA100Q25.
Panel 1-a shows the formation of a disk with a size of $\sim 20$ AU at $t_*=1.1 \times 10^{3}$ yr (where $t_*$ denotes the time after protostar formation).
At this stage, the total magnetic resistivity  is $\sim 10^{19} \etacm$ and relatively large (panel 1-b).
This large value is attributed to the fact that the maximum dust size remains $a_d<1 \mum$ within the disk (panel 1-c) and dust adsorption of charged particles is effective.
Owing to the efficient magnetic diffusion, the plasma $\beta$ inside the disk is significantly high (panel 1-d).

As the time progresses, a remarkable change in the magnetic resistivity occurs.
Panel 2-b shows a significant decrease in the magnetic resistivity within the disk.
This is caused by the dust growth within the disk ($ a_d$ reaches up to $10 \mum$) and reduction in the dust adsorption efficiency
(see Appendix B and \citet{2022ApJ...934...88T} for the impact of dust size on the ambipolar resistivity).
However, at this epoch, the decrease in the magnetic diffusion efficiency does not lead to a decrease in the disk size; instead, the disk continues to expand. The plasma $\beta$ also remains high.

The decrease in the magnetic resistivity leads to a better coupling between the magnetic field and gas in the disk.
This coupling promotes gas accretion, which in turn transports the magnetic flux to the center and reduces the plasma $\beta$ in the disk (panel 3-d).
The increase of magnetic field amplifies the magnetic resistivity  ($\eta_A$) in the central region (panel 3-b).
Once the dust grows sufficiently, $\eta_A$ is proportional to the square of the magnetic field strength even inside the disk.
Even at this point, the gas density map (panel 3-a) indicates the presence of a relatively large disk of $\sim 50$ AU.

The strong magnetic field in the disk causes efficient magnetic braking and efficient mass accretion over the entire disk, leading to a decrease in the disk size from 3-a to 4-a.
However, the density structure of the central region is very similar in panels 3-a and 4-a.
On the other hand, between panels 2-a and 4-a, the disk size is similar, but the density structure of the inner region is different.
This indicates that the inner disk structure transits from 2-a to 3-a as the dust grains grow.

In this way, the growth of the dust grains causes a decrease in the magnetic resistivity and changes the magnetic activity of the disk,
and ultimately determines the evolution of the disk.
Conversely, changes in the disk structures affect the dust growth in the disk (figure \ref{dust_size_mass}).
Based on these results, we propose a new evolutionary process of the protoplanetary disk: ``co-evolution of dust grains and protoplanetary disks".

\begin{figure*}
  \includegraphics[clip,trim=0mm 0mm 0mm 0mm,width=50mm,,angle=-90]{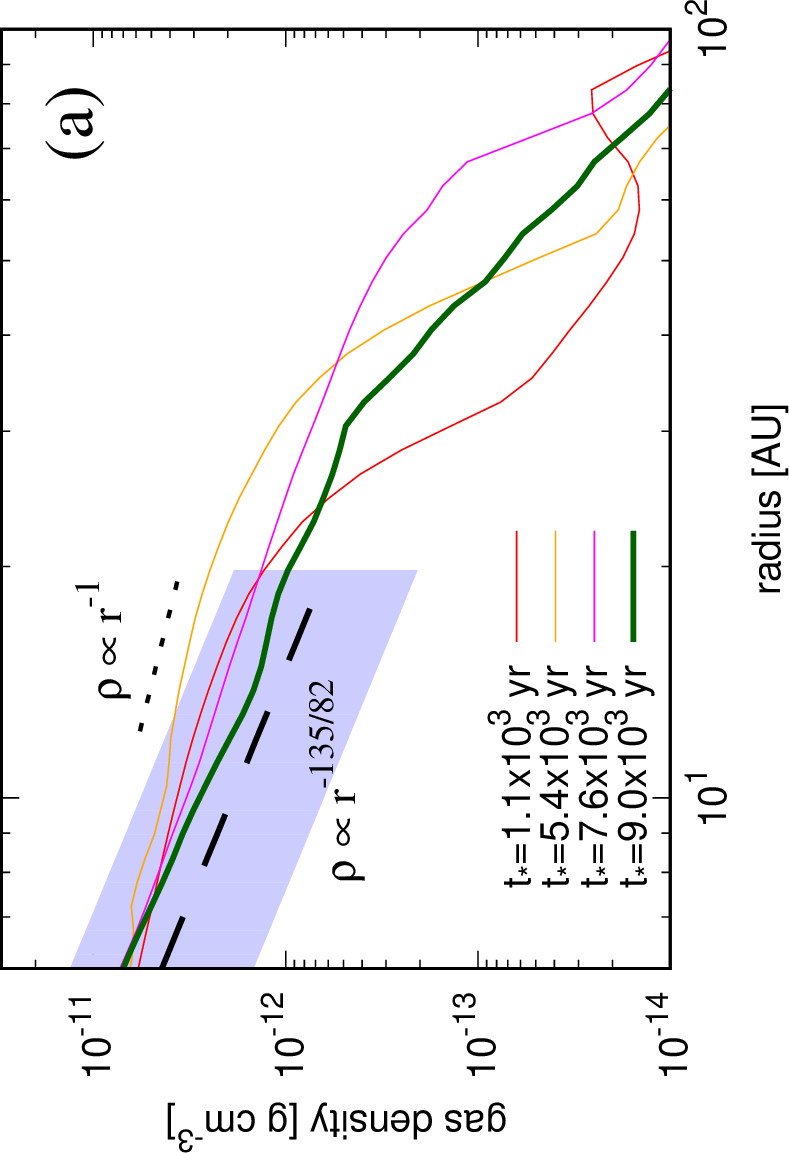}
  \includegraphics[clip,trim=0mm 0mm 0mm 0mm,width=50mm,,angle=-90]{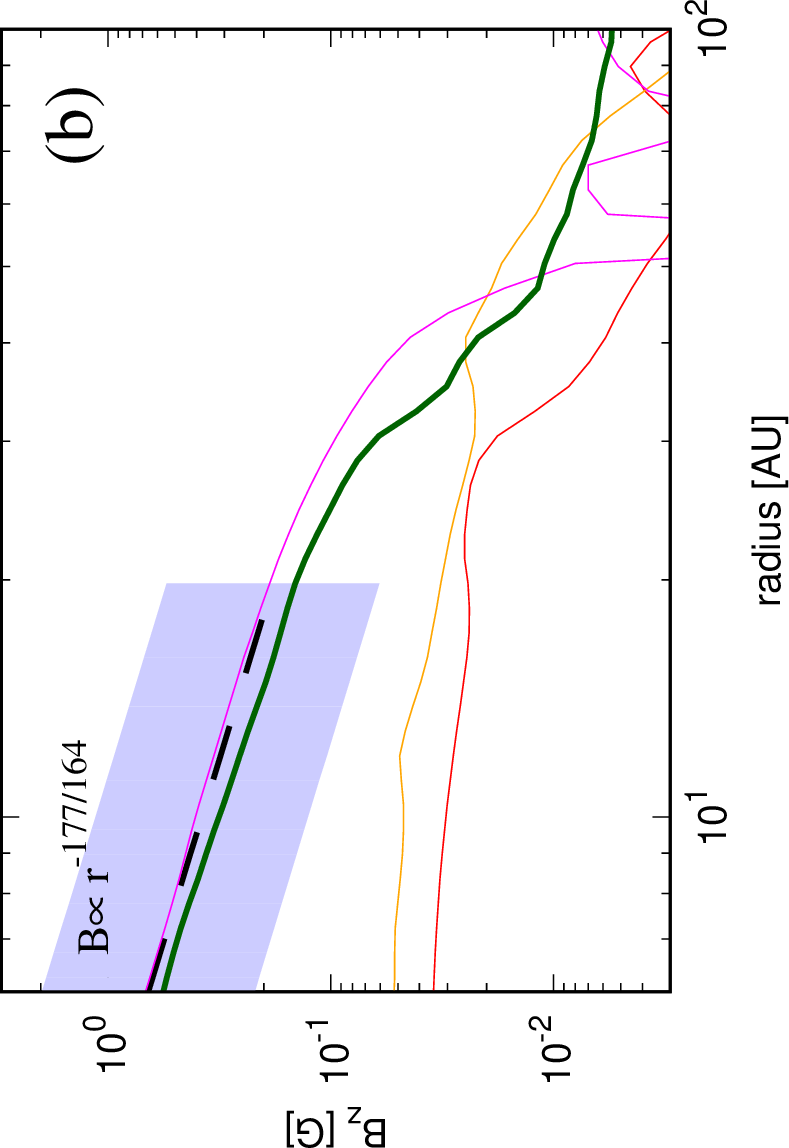}
  \includegraphics[clip,trim=0mm 0mm 0mm 0mm,width=50mm,,angle=-90]{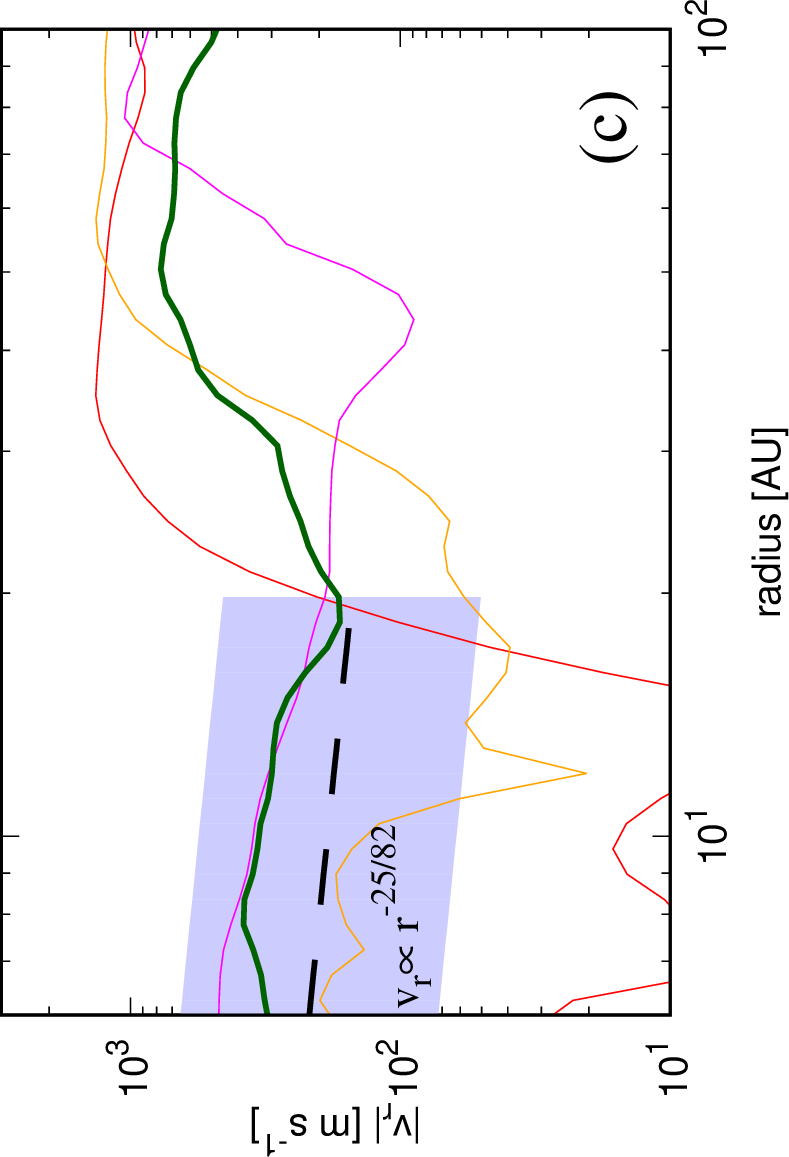}
  \includegraphics[clip,trim=0mm 0mm 0mm 0mm,width=50mm,,angle=-90]{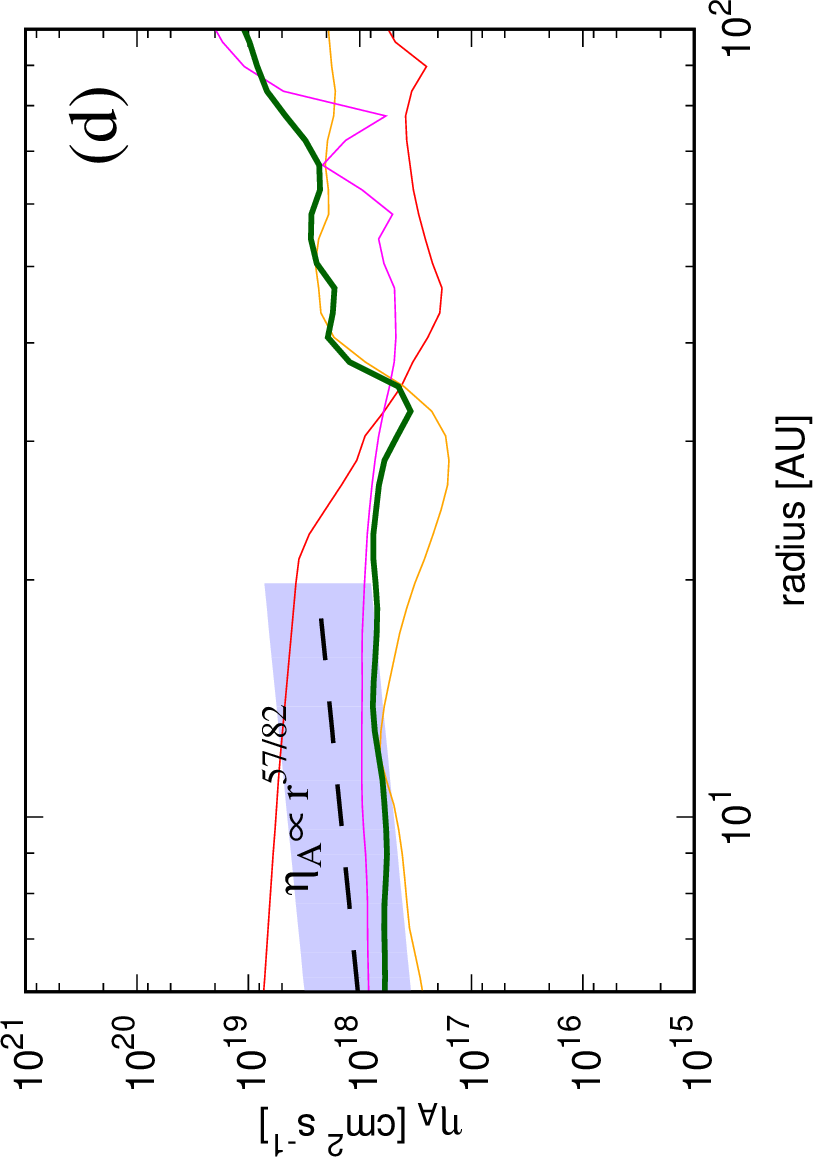}
  \caption{
    Time evolution of the azimuthally averaged radial profile of (panels a) gas density, (b) vertical magnetic field, (c) radial velocity,
    and  (d) ambipolar resistivity ($\eta_A$).
    The epoch of each line is the same as in figure \ref{2D_time_evolution_q25_a100nm}.
    The black dashed lines indicate the analytical solutions (equations (\ref{solution_first}) to (\ref{solution_last}))
    in which following values
    $\rho_c=4\times 10^{-14} \gcm$,
    $\zeta_{\rm CR}=10^{-17} {\rm s^{-1}}$,
    $c_{s, ref}=190 \ms$,
    $\dot{M}=2\times 10^{-5} \msunyear$, and  $M=0.3 \msun$  are assumed.
    The hatched areas are regions within a factor of three from the solutions.
  }
\label{1D_time_evolution_q25_a100nm}
\end{figure*}

\subsection{Radial disk structure and comparison with analytical solutions}
Figure \ref{1D_time_evolution_q25_a100nm} shows the azimuthally averaged radial profiles at the midplane of ModelA100Q25.
The red, orange, magenta, and green lines show profiles at the epochs of panels 1-4 in figure \ref{2D_time_evolution_q25_a100nm}, respectively.

In the early evolutionary phase when the dust size is sufficiently small ($t_* \lesssim 5\times 10^3$ yr; red and orange lines),
the gas density has a relatively shallow profile with power of  $D_{\rho_g} \sim -1$ (red and orange lines; where $D_f$ denotes the power exponent of a quantity $f$ as $f(r) \propto r^{D_f}$ ).
As the dust grains grow, the gas density profile becomes steeper and appears to converge to a power law with  $D_{\rho_g} \sim -2$ (at $t_*= 9\times 10^3$ yr; green line).

To explain this disk structure,
we analytically derive the new steady state solutions of the disk in Appendix A.
The important assumptions in deriving the steady state solutions are that
(1) the magnetic braking determines the disk angular momentum evolution (and angular momentum transfer by the viscosity is negligible), (2) radial magnetic flux transport is determined by the balance between gas advection and ambipolar diffusion, and (3) the adsorption of charged particles on the dust grains is negligible, and the ionization degree is determined by cosmic ray ionization and gas phase recombination ( for the details of the derivation, see Appendix A).

The derived solution of the density profile predicts $D_{\rho_g} = -\frac{135}{82}$ (equation (\ref{solution_first})). 
The black dotted line and hatched areas indicate the analytical solutions and a region within a factor of {\rm three} of the solution, respectively.
Here, we have chosen the following values for the analytical solution:
$\rho_c=4\times 10^{-14} \gcm$,
$\zeta_{\rm CR}=10^{-17}  {\rm s^{-1}}$,
$c_{s, ref}=190 \ms$,
which are led from the simulation setup and 
$\dot{M}=2\times 10^{-5} \msunyear$ and  $M=0.3 \msun$ to be approximately consistent with the values at $t_* = 9 \times 10^3$ yr.
Our analytical solution well agrees with the simulation result not only in terms of the power, but also in terms of the exact value.

Panel (b) shows that in the early evolutionary phase (red and orange lines),
the magnetic field in the disk is almost constant ($D_{B_z} \sim 0$), which is in agreement with previous studies \citep{2016A&A...587A..32M,2015MNRAS.452..278T} in which the dust growth is ignored.
In contrast, as the dust grains grow, the vertical magnetic field profile becomes steeper and converges to a power law with $D_{B_z} \sim -1$.
Our analytical solution suggests the power exponent of $D_{B_z} = -\frac{177}{164}$ (equation (\ref{solution_second})) and agree with the simulation results at $t_*=9.0 \times 10^3$ yr (green line).
The black dotted line indicates our analytical solution with the same parameters used in the density profile and the hatched area are a region within a factor of  three from the solution.
The black dotted line confirms that our analytical solution simultaneously reproduces the density and magnetic field with a single set of parameters and quantitative agrees with the simulation results.

Panel (c) shows that, when the dust size is small, the absolute value of radial velocity is  $ |v_r|\lesssim 10 \ms$ (red line).
As the dust grows, it increases in the order of $100 \ms$.
The proposed analytical solution suggests a power of $D_{v_r} = -\frac{25}{82}$ and value of $\sim 100 \ms$. Although $v_r$ possesses relatively strong time fluctuation, the simulation results at $t_*=9.0 \times 10^3$ yr (green line) agrees with the analytical solutions.

Panel (d) shows that the $\eta_A$ profile in the simulation is almost radially constant once the dust grains sufficiently grow (magenta and green lines).
On the other hand, our analytical model predicts slightly positive power law with $D_{\eta_A}=\frac{57}{82}$.
Compared to other physical quantities, the difference between the simulation result and the analytical solution is relatively large, but still within a factor of three
in region $r \lesssim 10$ AU.


\subsection{Diversity and universality of disk evolution}
As  seen in the previous section, the disk evolution is significantly affected by the dust growth.
Furthermore, once the dust grains sufficiently grows,
the disk structure of our fiducial model is well described by the power laws analytically derived in Appendix A.
In this section, we examine the impact of the minimum dust size and dust power exponent on the evolution of the disk.
Moreover we examine the effect of the cosmic-ray ionization rate.
Through these considerations, we discuss the diversity and universality of protoplanetary disk evolution.

Before discussing the simulation results, we summarize how the minimum dust size $a_{\rm min}$ and the power exponent $q$ affect the magnetic resistivity based on our previous studies \citep{2022ApJ...934...88T}.
  Figures which show how the resistivities depend on $a_{\rm max}$ with various $q$ and $a_{\rm min}$ can be found in \citet{2022ApJ...934...88T}.

The adsorption of dust grains, which plays the primary role in determining the resistivity depends on the dust total cross-section, which depends on $q$ and $a_{\rm min}$.
In the case where $q$ is less than 3, the maximum dust size determines the total surface area. Conversely, in a case where $q$ is larger than 3, both the maximum and minimum dust sizes affect the total surface area. Consequently, when $q$ is large (or the size distribution is steep), the influence of dust growth tends to be weaker.

Another important factor that influences resistivity is conductivity generated by the dust grains.
If a significant amount of dust with a size of  $\lesssim 10 \nm$ is present, the dust grains contributes to conductivity. In such case, resistivity at a high density
is smaller than that when the minimum size is, for instance, $100 ~\nm$.
This effect is pronounced in a case when the dust grains have not grown and the size distribution is steep (i.e., $q$ is large).

\subsubsection{Comparison of density structures}
\begin{figure*}
    \includegraphics[clip,trim=0mm 0mm 0mm 0mm,width=150mm]{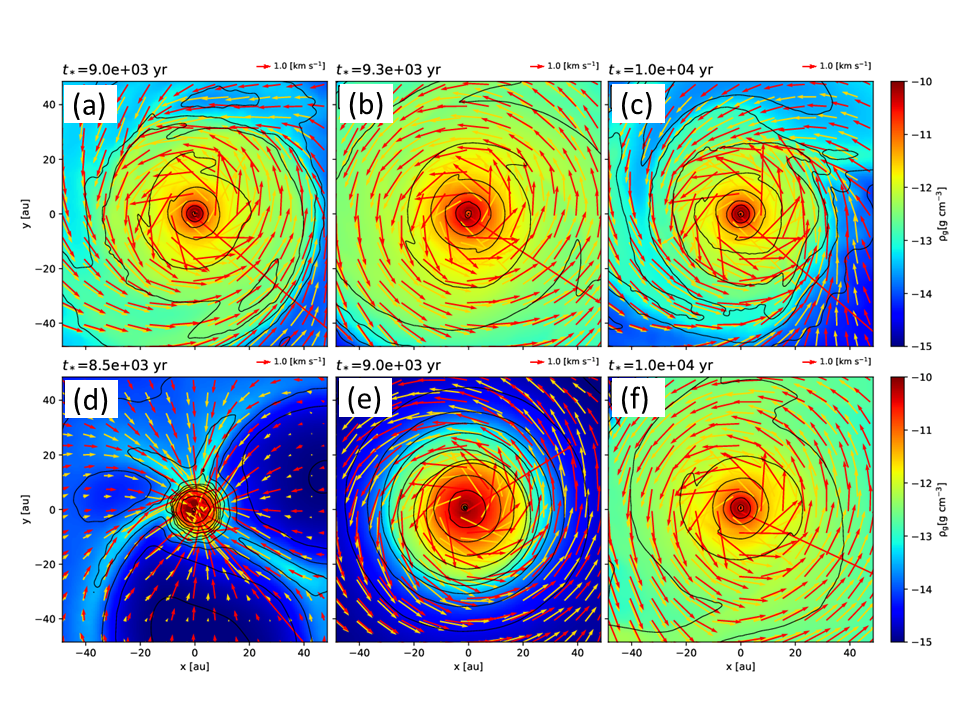}
  \caption{
    Gas density map of (a) ModelA100Q25, (b) ModelA100Q35, (c) ModelA5Q25, (d) ModelA5Q35, (e) ModelA100Fixed, (f) ModelZeta18 on the $x$-$y$ plane.
    The time after protostar formation are shown in upper left.
    The red and orange arrows show the velocity field of the dust and gas, respectively, on the $x$-$y$ plane.
    The black lines indicate the density contour.
    The contour levels are $\rho_g=10^{-15},10^{-14.5},\cdots,10^{-10}\gcm$.
}
\label{2D_density_all_models_end}
\end{figure*}

Figure \ref{2D_density_all_models_end} shows the density map of all models.
Although the disk size is different among the simulation, the density structure of inner $\sim 20$ AU region of panels a (ModelA100Q25), b (ModelA100Q35), c (ModelA5Q25), and f (ModelZeta18) are very similar.
In these simulations, the inner density structures are consistent with the analytical solutions.
The spiral patterns in the outer regions of the disks are created by gravitational instability. We confirm that Toomre's $Q$ parameter in these regions are $Q \sim 1$ in the outer regions of these disks.

On the other hand, figure \ref{2D_density_all_models_end} d (ModelA5Q35) shows the formation of a very small disk of $\lesssim 10 {\rm AU}$ and bubble-like structures around it.
This bubble-like structure is created by magnetic interchange instability \citep{2012ApJ...757...77K}.
The large amounts of small dust grains make ambipolar diffusion (and Ohmic diffusion) ineffective in the high-density region and leads to the development of interchange instability as a redistribution mechanism of magnetic flux.

Figure \ref{2D_density_all_models_end} e (ModelA100Fixed) in which the dust growth is artificially ignored shows that the disk is relatively compact, dense and massive.
This massive disk is consistent with previous theoretical studies; however such massive disk seems to be inconsistent with the observations \citep{2022arXiv220913765T}.

\subsubsection{Universality of the disk structure}
\begin{figure*}
  \includegraphics[clip,trim=0mm 0mm 0mm 0mm,width=50mm,,angle=-90]{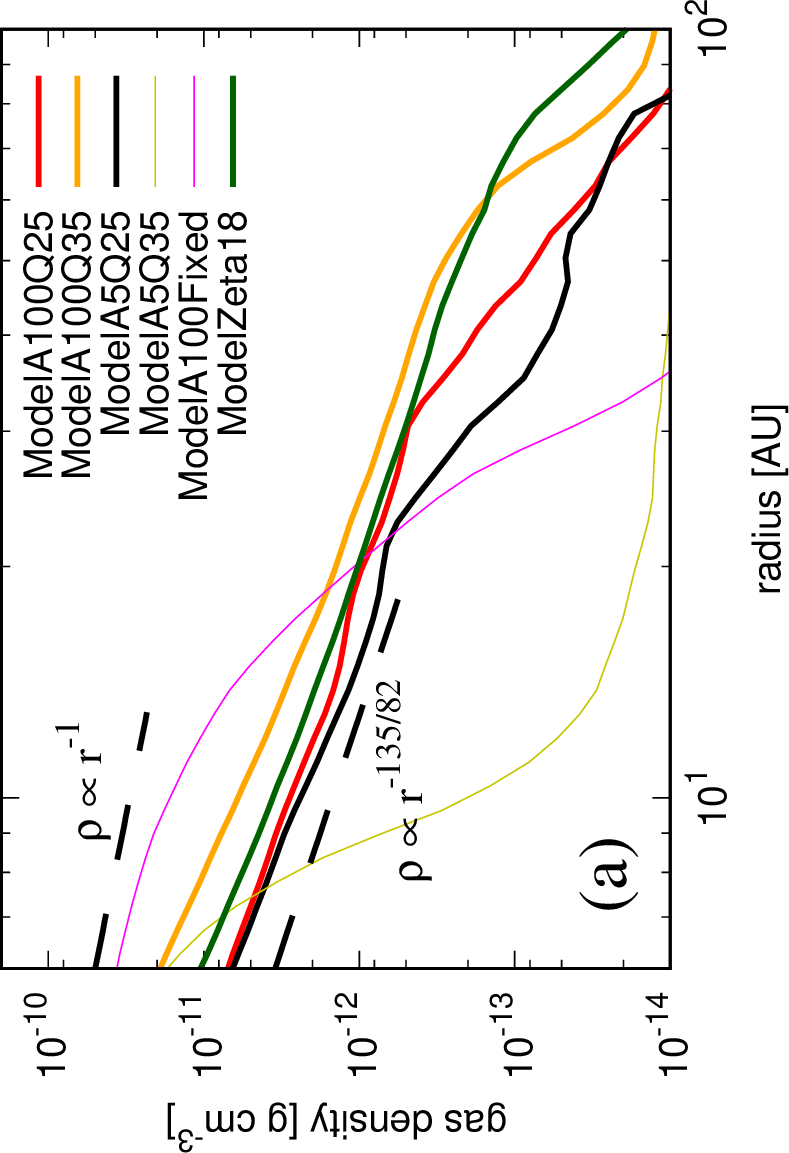}
  \includegraphics[clip,trim=0mm 0mm 0mm 0mm,width=50mm,,angle=-90]{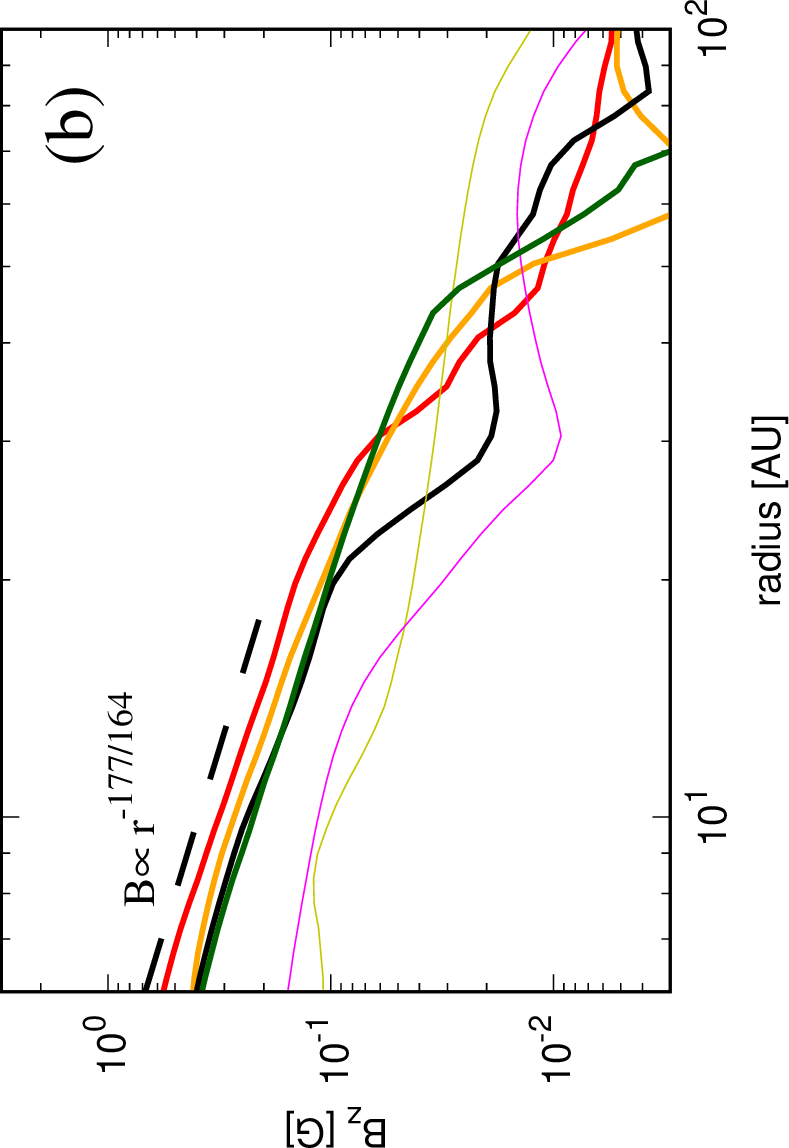}
  \includegraphics[clip,trim=0mm 0mm 0mm 0mm,width=50mm,,angle=-90]{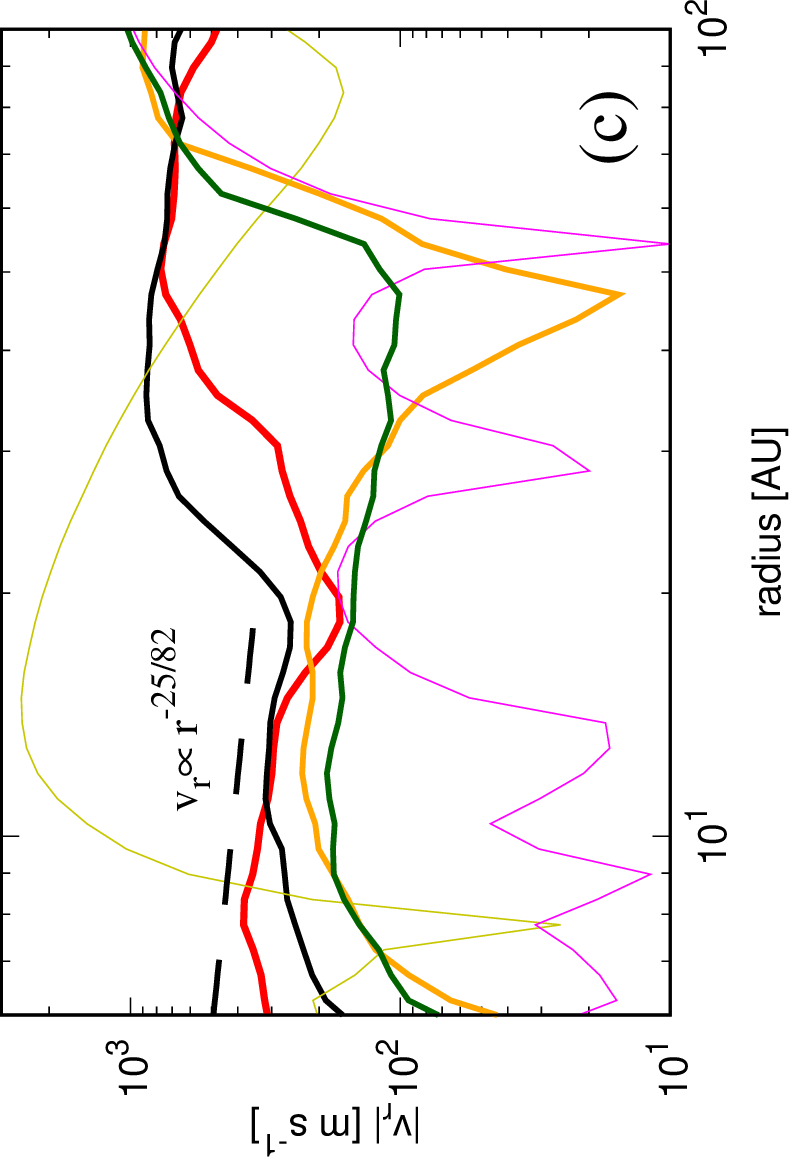}
  \includegraphics[clip,trim=0mm 0mm 0mm 0mm,width=50mm,,angle=-90]{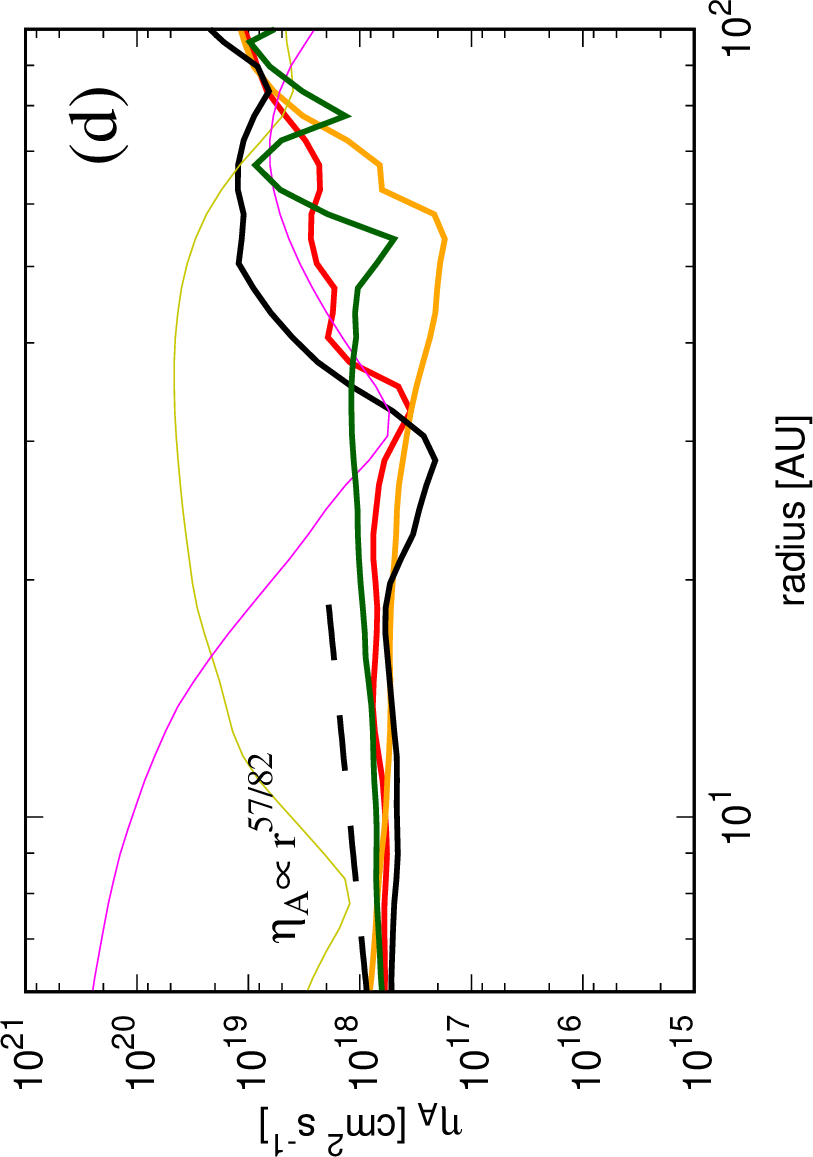}
  \caption{
    The azimuthally averaged radial profile of (a) gas density, (b) vertical magnetic field, (c) radial velocity,  and  (d) ambipolar resistivity ($\eta_A$).
    The red, orange, black, yellow, magenta, and green lines
    show the results of ModelA100Q25, ModelA100Q35, ModelA5Q25, ModelA5Q35, ModelA100Fixed, ModelZeta18, respectively.
    The thick lines show the models that are in good agreement with the power laws of the analytical solution.
    The time of each line is the same as in figure \ref{2D_density_all_models_end}.
    The black dashed lines indicate the power laws of the analytical solutions.
}
\label{rad_prof_all_models}
\end{figure*}

Figure \ref{rad_prof_all_models} shows the azimuthally averaged radial profiles of the all models.
The dashed lines show the power laws of our analytical solutions (in this figure, we plot the power laws just as a reference because parameters such as central star mass or mass accretion rate differ among the models). 

Of the five models that consider dust growth, four models have an inner region of $\sim 20 {\rm AU}$ consistent with the analytical solution (red, green, black, and orange lines).
If we regard the disk radius as the radius at which the density distribution deviates from the power law of the analytical solution, ModelZeta18 has the largest disk size, and ModelA100Q35, ModelA100Q25, and ModelA5Q25 have smaller disk sizes in this order (see also figure \ref{disk_size}).
This difference may be due to the difference in resistivity in the regions where dust grains have not grown (outer region of the disk and envelope). Despite the difference in disk size, the universality of the disk structure in the inner region is noteworthy.

Note that the disk structure of  ModelZeta18 (green lines) is
more consistent with the power laws of the analytical solutions than the fiducial model. For example, $\eta_A$ has a positive power exponent. This is because this model is more evolved than the fiducial model and approximation on $\eta_A$ in the analytical solution is better validated.

Figure \ref{rad_prof_all_models} shows that the disk of ModelA5Q25 (yellow) is very small and has a different structure from the other four models.
A large amount of small dust in this model makes ambipolar diffusion ineffective from the beginning of disk formation in the high-density region.  The disk rapidly shrinks before the dust grows and $\eta_A$ is well described by the analytical form of Shu \citep{1983ApJ...273..202S}. Thus, the disk evolution of the model is different from other models.

It would be instructive to see the differences in disk structure between the model ignoring the dust growth (ModelA100Fixed; magenta) and those that well described by the analytical solutions.
In the model without the dust growth, the density and $\eta_A$ are large, and the magnetic field and radial velocity $v_r$ are small.
This is because in the absence of the dust growth, the ambipolar diffusion in the disk is extremely effective.
It suppresses magnetic braking in the disk, resulting in smaller $v_r$ and an increased density due to gas accumulation in the disk.
Furthermore, the magnetic flux is extracted from the disk by the ambipolar diffusion causing the low value of the disk magnetic field.
The density of the disk in ModelA100Fixed is large and Toomre's $Q$ parameter is $Q\sim 1$  even in the inner region.


\subsection{Time evolution}
In this section, we examine the time evolution of disk angular momentum (and radius), disk  mass, typical dust size and dust abundance in disks. We also investigate the mass ejection rate by outflows from the disks.

\subsubsection{Time evolution of disk size}
\begin{figure*}
  \includegraphics[clip,trim=0mm 0mm 0mm 0mm,width=50mm,,angle=-90]{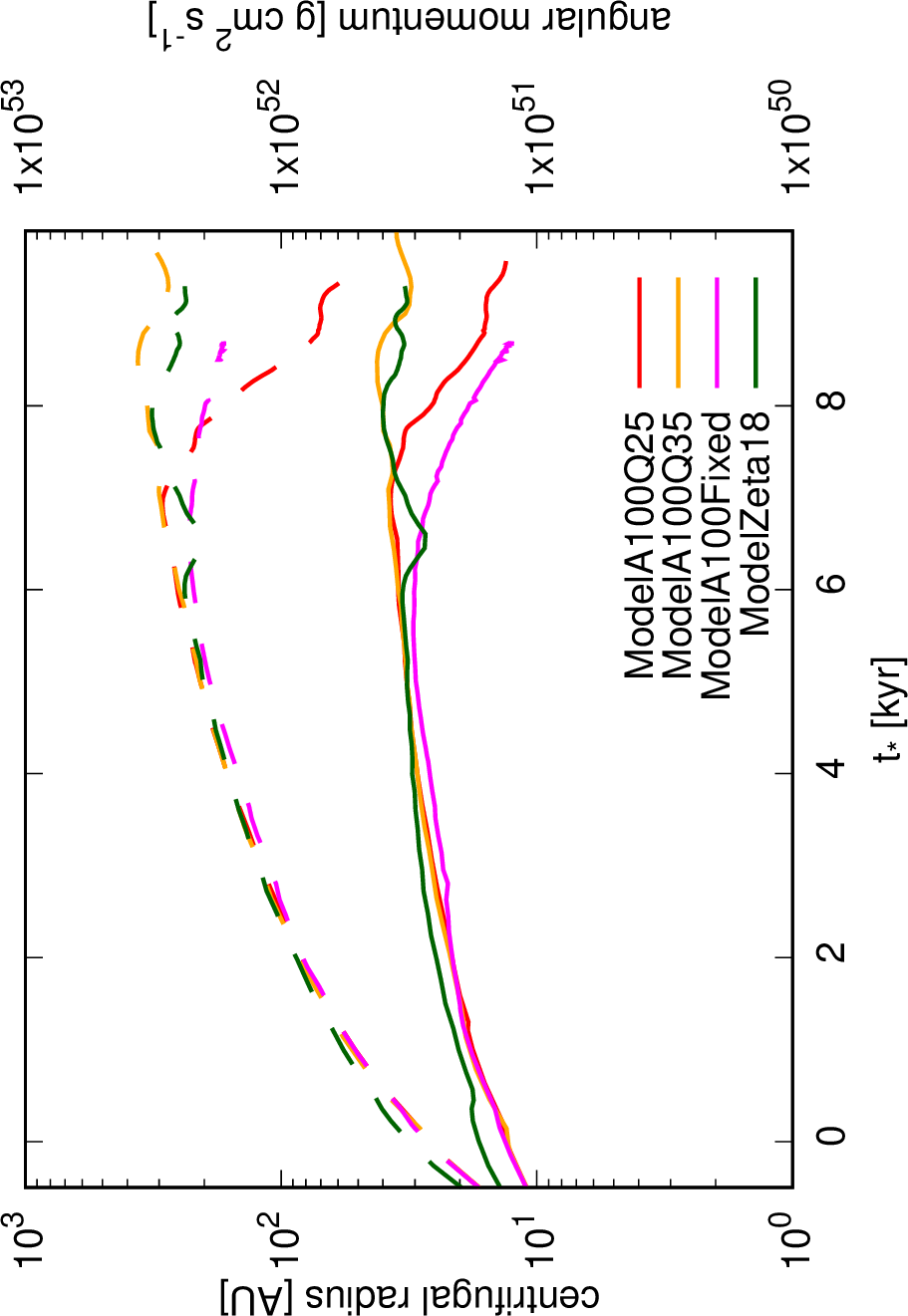}
  \includegraphics[clip,trim=0mm 0mm 0mm 0mm,width=50mm,,angle=-90]{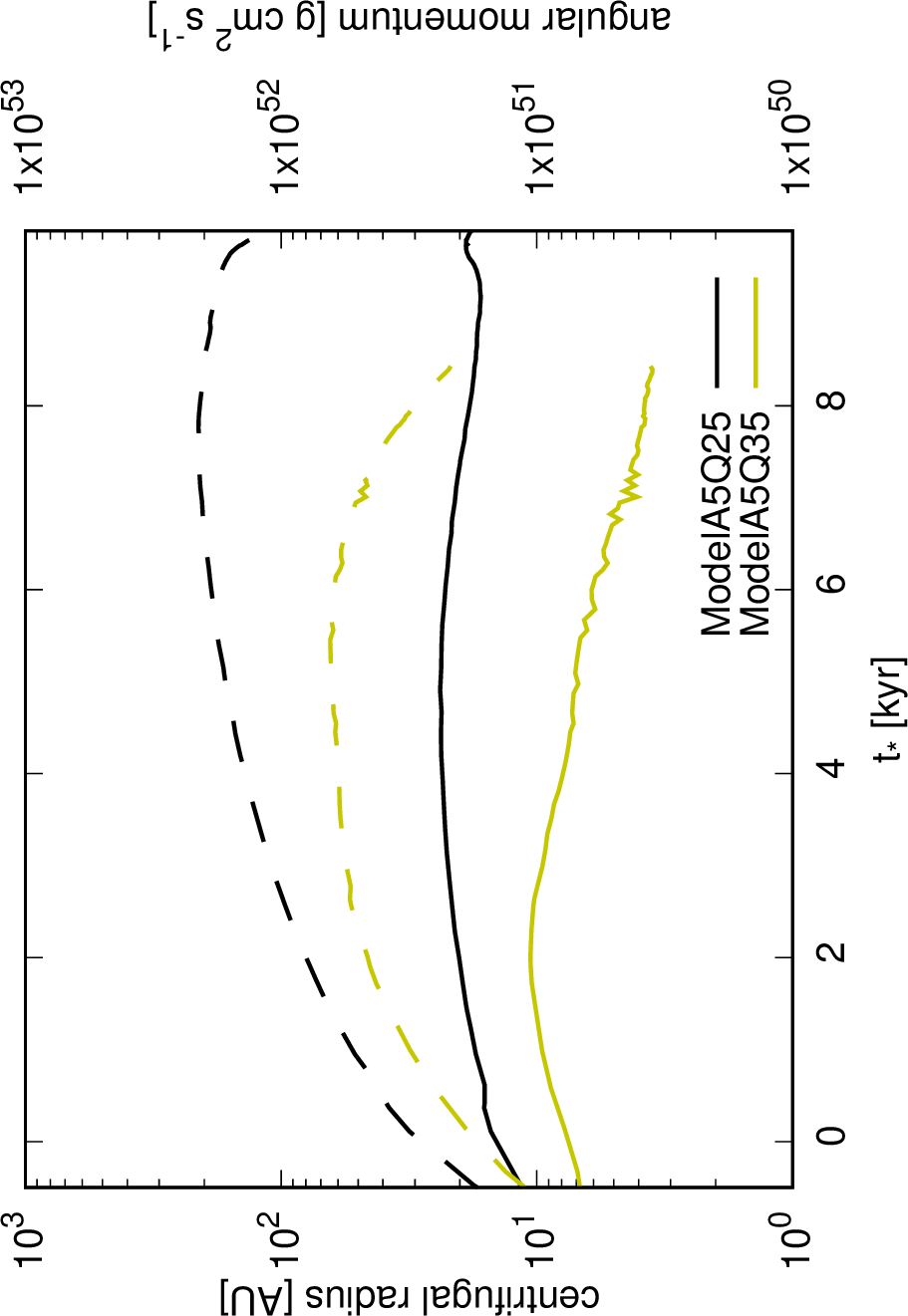}
  \caption{
    Time evolution of centrifugal radius (the solid lines and the left axis) and total angular momentum of disk (the dashed lines and the right axis).
    Red, orange, black, yellow, magenta, and green lines
    show the results of ModelA100Q25,ModelA100Q35, ModelA5Q25, ModelA5Q35, ModelA100Fixed, ModelZeta18, respectively.
    The horizontal axis shows the time after the protostar formation.
}
\label{disk_size}
\end{figure*}

Figure \ref{disk_size} shows the time evolution of
centrifugal radius and the angular momentum of the disk.
The angular momentum of disk $J(\rho_{\rm disk})$ is calculated as
\begin{eqnarray}
J(\rho_{\rm disk})\equiv \left| \int_{\rho_g>\rho_{\rm disk}} \rho_g (\rad \times \vel) d V \right|.
\end{eqnarray}
For the density threshold of the disk, we choose $\rho_{\rm disk}=10^{-13} \gcm$.
The centrifugal radius is calculated as
\begin{eqnarray}
r_{\rm disk} \equiv  r_{\rm cent}=  \frac{\bar{j}(\rho_{\rm disk})^2}{G M_{\rm star}}.
\end{eqnarray}
Here $\bar{j}(\rho_{\rm disk})=J(\rho_{\rm disk})/M_{\rm disk}$, where $M_{\rm disk}$ denotes the enclosed gas mass within the region $\rho_g>\rho_{\rm disk}$.
Comparing the radius of the region with a density of $\rho_g > \rho_{\rm disk}$   (figure \ref{rad_prof_all_models}) with the centrifugal radius, the former is about 2 times larger than the latter owing to radial density distribution and temporal density oscillation by the non-axisymmetric structures. In this study, we consider the centrifugal radius as an estimate of the disk size.

The left panel of figure \ref{disk_size} shows the results of the models with $a_{\rm min}=100~\nm$. 
In ModelA100Q25 (red line),  the  angular momentum of the disk continues to increase until $t_* \sim 7\times 10^3$ yr, and then it shows a sharp decrease.
This is because the dust grains grows to $a_{\rm d}\gtrsim 10 \mum$ (figure \ref{dust_size_mass}), which causes a sudden decrease in magnetic resistivity and the extraction of angular momentum by magnetic field.
Interestingly, although the angular momentum has decreased by a factor of $\sim 1/3$ from $t_* \sim 7\times 10^3$ yr to $t_* \sim 9 \times 10^3$ yr, the centrifugal radius has decreased by only a factor of $\sim 1/2$.
This indicates that the disk mass also decreases rapidly during this period (figure \ref{disk_mass}).
For ModelA100Q35, no such rapid decrease of angular momentum is observed.
This is because $a_{\rm min}$  is also responsible for the total dust surface area, and thus the decrease in magnetic resistivity is not so drastic \citep{2022ApJ...934...88T}.
As exhibited by  ModelZeta18 (green line), the low cosmic-ray ionization rate can contribute to maintaining the angular momentum.
This is in agreement with previous studies \citep{2018MNRAS.476.2063W,2020A&A...639A..86K, 2023MNRAS.tmp..691K}.
The decrease in the disk size of ModelA100Fixed is caused by the pseudo-disk warp and associated inward magnetic flux drag \citep{2020ApJ...896..158T}.
In this model, the angular momentum decreases less than a factor of two, which is not significant compared to ModelA100Q25.

The right panel of figure \ref{disk_size} shows the results of the models with $a_{\rm min}=5~\nm$. 
Interestingly, the relationship between the disk size and power exponent $q$ is different from the models with $a_{\rm min}=100 \nm$.
The disk size of the model with $q=3.5$ (ModelA5Q35) is significantly smaller
than that in the model with $q=2.5$ (ModelA5Q25).
This is because when there is a large amount of small dust ($\sim \nm$), dust grains are responsible for the conductivity and reduce the magnetic resistivity.
This allows magnetic braking to work more effectively in ModelA5Q35 and reduce the disk size.

\subsubsection{Time evolution of the disk mass}
\begin{figure*}
  \includegraphics[clip,trim=0mm 0mm 0mm 0mm,width=50mm,,angle=-90]{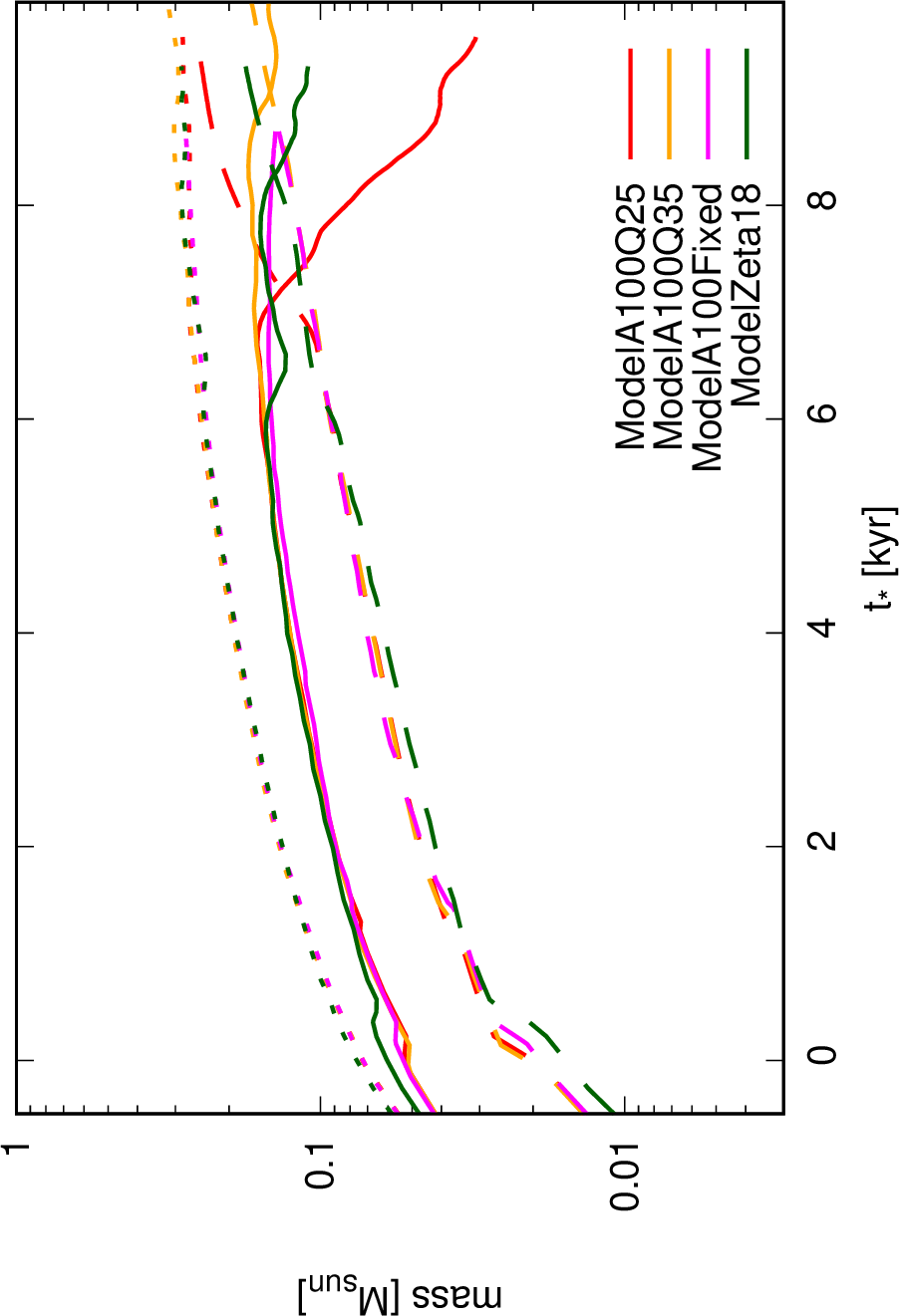}
  \includegraphics[clip,trim=0mm 0mm 0mm 0mm,width=50mm,,angle=-90]{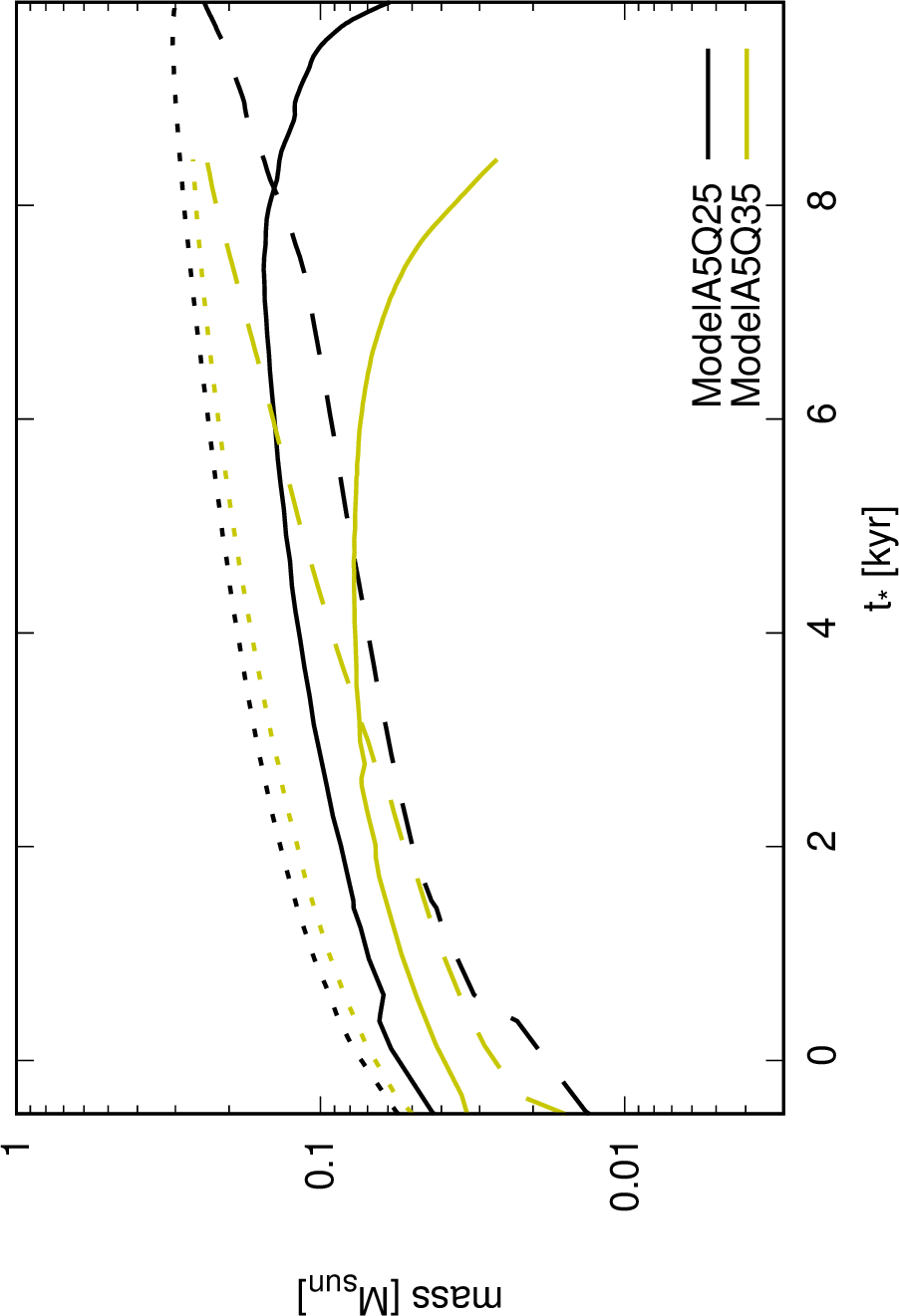}
  \caption{
    Time evolution of the mass of the disks (the solid lines), the protostars (the dashed lines), and the total mass (the dotted lines).
    Red, orange, black, yellow, magenta, and green lines
    show the results of ModelA100Q25, ModelA100Q35, ModelA5Q25, ModelA5Q35, ModelA100Fixed, ModelZeta18, respectively.
    The horizontal axis shows the time after the protostar formation.
}
\label{disk_mass}
\end{figure*}

Figure \ref{disk_mass} shows the time evolution of disk mass $M_{\rm disk}$ (solid), protostar mass $M_{\rm star}$ (dashed), and total mass $M_{\rm star}+M_{\rm disk}$ (dotted). 
In ModelA100Q25 (red), the disk mass continues to increase and reaches $\sim 0.15 \msun$ at $t_* \sim 7 \times 10^3$ yr. Then, it drops sharply to $M_{\rm disk}\sim 0.03 \msun$ at $t_* \sim 9 \times 10^3$ yr.
Meanwhile, mass accretion onto protostars is enhanced and the protostellar mass rapidly increases from $0.1 \msun$ to 0.2 $\msun$ in this period, giving a mass accretion rate of $\gtrsim 10^{-5} \msunyear$ for this model.
In ModelZeta18 (green line) and ModelA5Q25 (black line), it appears that the disk mass also begins to decrease towards the end of the simulations. However, the time at which the decrease begins is later than in ModelA100Q25.

In ModelA100Fixed, the disk radius decreases in $t_*\gtrsim 6\times 10^3$ yr (figure \ref{disk_size}) but the disk mass does not significantly change.
This suggests that the disk evolution without dust growth is different from the models that include dust growth and that are consistent consistent with the analytic solution.
In ModelA5Q35, the mass (and radius) evolution differs from the other models.
This is due to inefficient ambipolar diffusion since disk formation. Thus, the situation close to the ideal MHD is realized.

\subsubsection{time evolution of dust size and dust abundance in the disk}

Figure \ref{dust_size_mass} shows the time evolution of the dust-to-gas mass ratio and mean dust size of the disks.
The dust mass and mean dust size of the disk is calculated as
\begin{eqnarray}
\bar{a}_{\rm disk}\equiv \frac{1}{M_{\rm disk}}\int_{\rho_g>\rho_{\rm disk}} \rho_g a_d d V,
\end{eqnarray}
and 
\begin{eqnarray}
M_{\rm dust, disk} \equiv \int_{\rho_g>\rho_{\rm disk}} \rho_d d V,
\end{eqnarray}
respectively.

The figure shows that the increase in the dust-to-gas mass ratio occurs later in the simulation.
This increase begins when the average dust size in the disk exceeds $\sim 100 \mum$.
This is due to the ``ash-fall phenomenon" proposed in our previous study \citep{2021ApJ...920L..35T}.
The largest increase is observed in ModelA100Q25, where the dust-to-gas mass ratio increases to 1.04\% at the end of the simulation.
Some readers may think that this value is small and irrelevant.
However, we only considered $\sim 10^3$ yr after the ratio started to increase.
If this event continues for, for instance, $10^5$ yr (i.e., during the Class 0/I phase), it can cause a significant increase in the dust abundance.

The increase of the dust-to-gas mass ratio is slower in ModelA100Q35, ModelA5Q25, and ModelZeta18, compared to ModelA100Q25 owing to the lower outflow activity (i.e., weaker coupling between the magnetic field and gas in the upper layers of the disk;  figure \ref{Fout}) in these models.

\subsubsection{Time evolution of the mass ejection rate by outflow}
\begin{figure*}
  \includegraphics[clip,trim=0mm 0mm 0mm 0mm,width=50mm,,angle=-90]{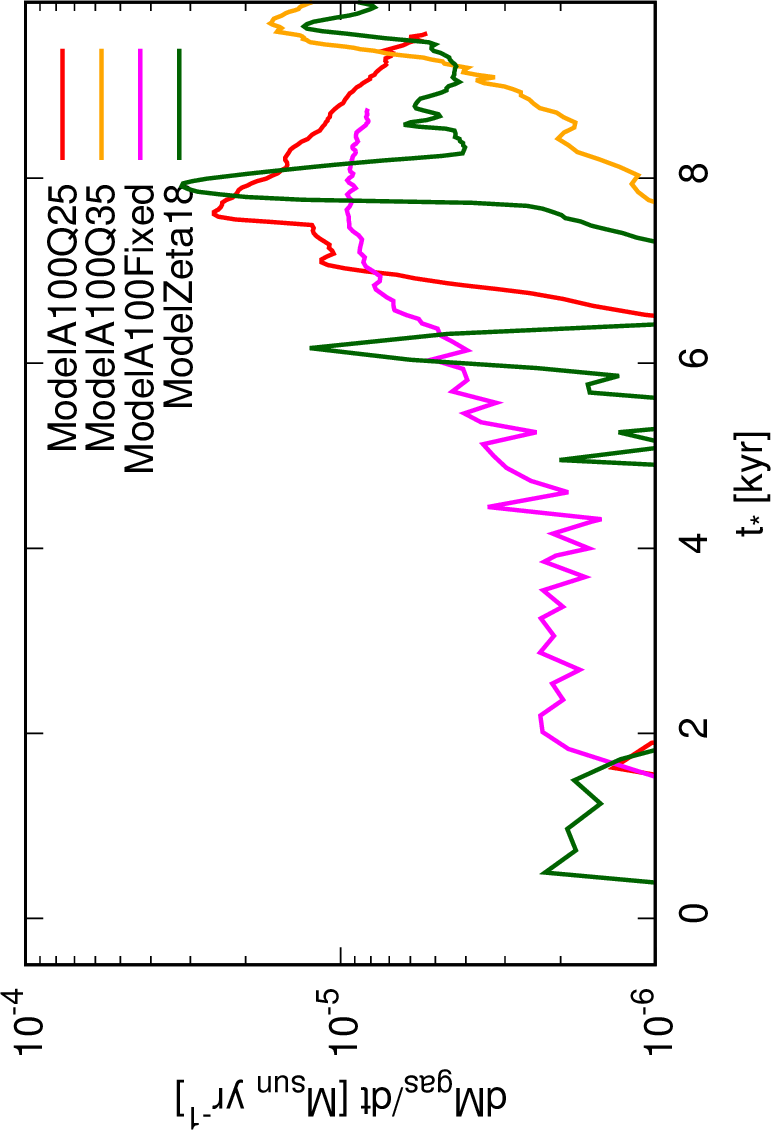}
  \includegraphics[clip,trim=0mm 0mm 0mm 0mm,width=50mm,,angle=-90]{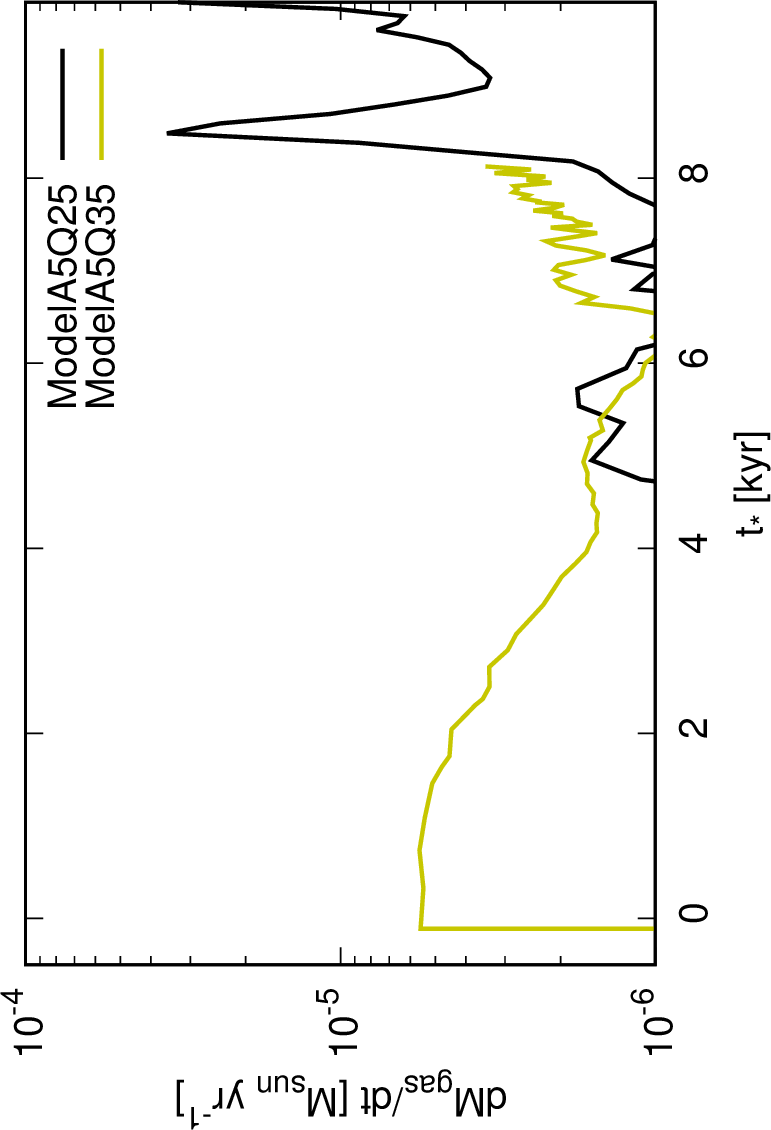}
  \caption{
    Time evolution of the mass ejection rate.
    Red, orange, black, yellow, magenta, and green lines
    show the results of ModelA100Q25, ModelA100Q35, ModelA5Q25, ModelA5Q35, ModelA100Fixed, ModelZeta18, respectively.
    The horizontal axis shows the time after the protostar formation.
}
\label{Fout}
\end{figure*}

Figure \ref{Fout} shows the time evolution of the mass ejection rate due to the molecular outflow.
The mass ejection rate is calculated as
\begin{eqnarray}
\dot{M}_{\rm out} \equiv\int_{r=100 {\rm AU}} \rho v^+_r dS,
\end{eqnarray}
where $v_r^+ \equiv \max(\vel\cdot \rad,0)$ denotes the positive radial velocity, and we perform the surface integral on a sphere with radius of $100 {\rm AU}$.

\begin{figure*}
  \includegraphics[clip,trim=0mm 0mm 0mm 0mm,width=50mm,,angle=-90]{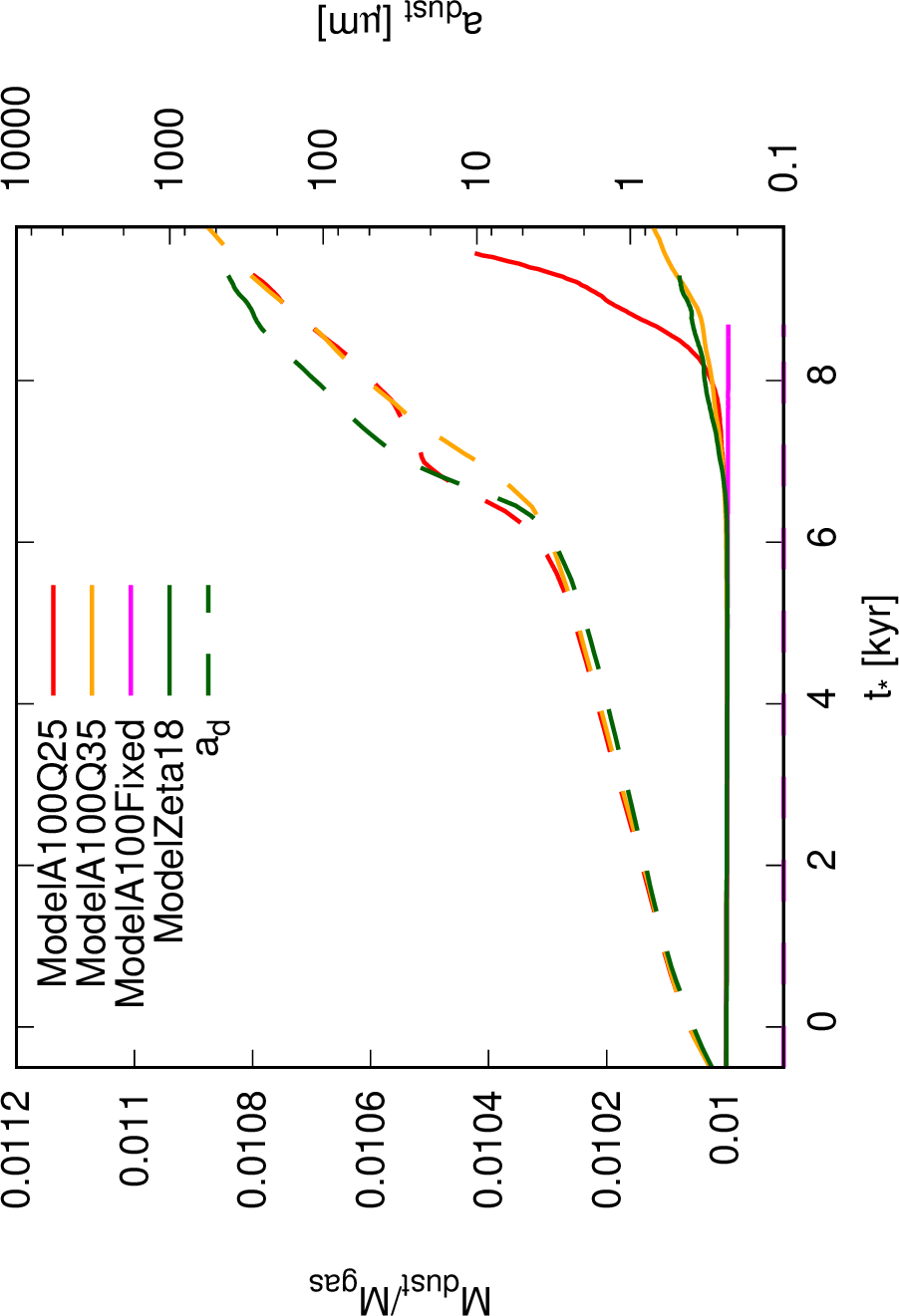}
  \includegraphics[clip,trim=0mm 0mm 0mm 0mm,width=50mm,,angle=-90]{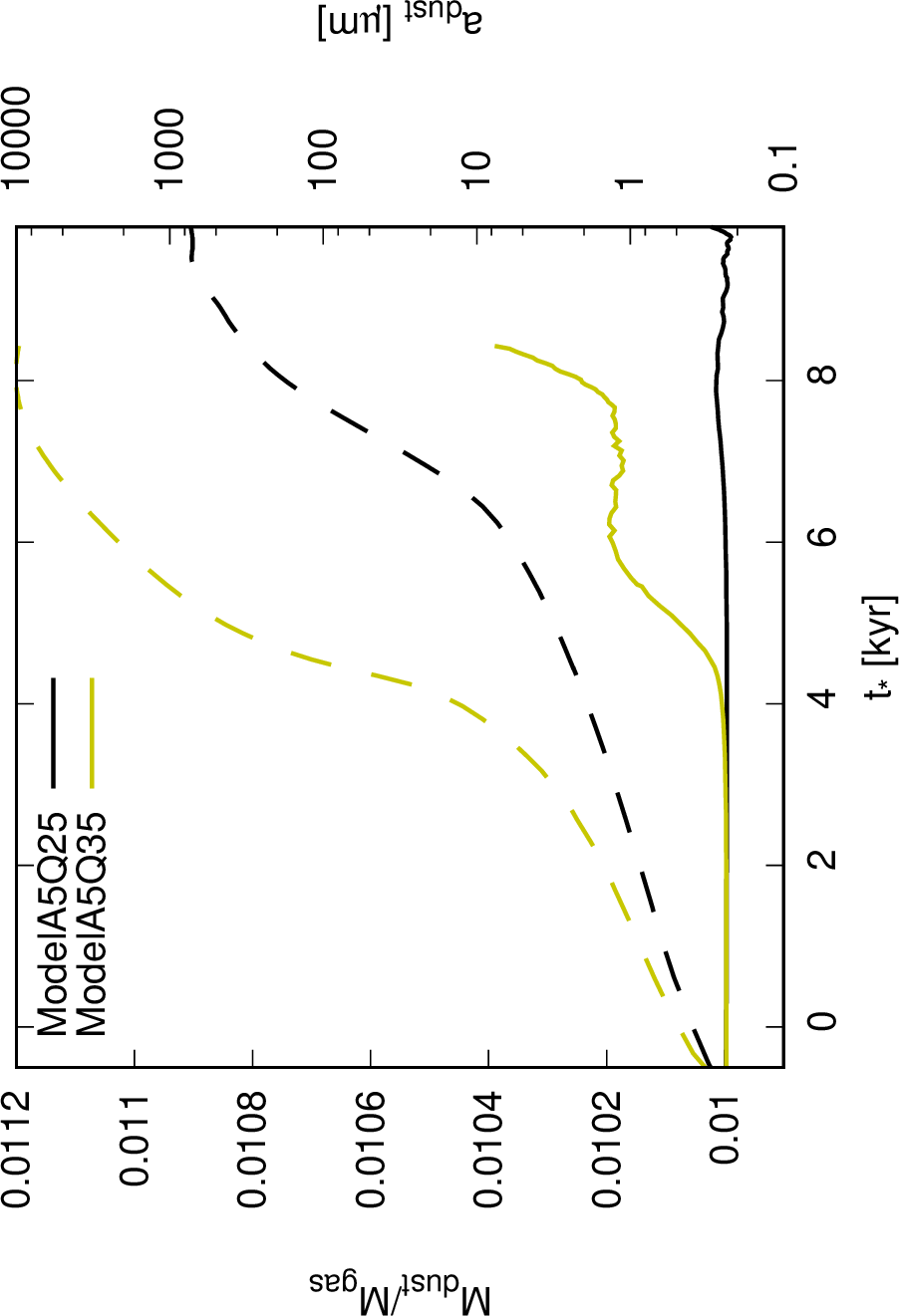}
  \caption{
    Time evolution of the dust-to-gas mass ratio (the solid lines and left axis) and  mean dust size in the disk  (the dashed lines and right axis).
    Red, orange, black, yellow, magenta, and green lines
    show the results of ModelA100Q25, ModelA100Q35, ModelA5Q25, ModelA5Q35, ModelA100Fixed, ModelZeta18, respectively.
    The horizontal axis shows the time after the protostar formation.
}
\label{dust_size_mass}
\end{figure*}

The mass ejection rate is highly variable and has a peak value of $\gtrsim 10^{-5} \msunyear$. This is comparable to the mass accretion rate in the disk.
The figure \ref{dust_size_mass} and \ref{Fout} suggest that a mass ejection rate of $10^{-6}$ to $10^{-5} \msunyear$ and the mean dust size of $\sim 100 \mum$ are required for the "ash-fall" phenomenon to happen, in which the disk gas is selectively ejected into interstellar space by the outflow and the dust grains are resupplied to the disk.

\section{Discussion}
\subsection{Universality of the disk structure in the disk with grown dust grains}
In this study, we propose the new disk evolutionary picture ``co-evolution of dust grains and protoplanetary disks" based on non-ideal dust-gas two-fluid MHD simulations considering dust growth. The dust growth changes the gas-phase ionization degree, magnetic resistivity, and evolution of the disk.
In the co-evolution process, the microscopic dust grains of micrometer size couple with the macroscopic disks of $100$ AU size and they co-evolve. The size scale difference between two objects is $10^{19}$, which is astounding compared to the well-known co-evolution of super massive black holes and galaxies (size scale difference is $\sim 10^{10}$).

Furthermore, once the dust grains grow sufficiently, the structure of protoplanetary disks is well described by the non-trivial power laws, which we analytically derive in equations (\ref{solution_first}) to  (\ref{solution_last}) (figure \ref{1D_time_evolution_q25_a100nm} and \ref{rad_prof_all_models}).
From the assumptions adopted in the analytical solutions, we conclude that the disk structures will emerge when
\begin{enumerate}
\item Dust grains grow sufficiently and adsorption of charged particles by the dust grains becomes negligible,
\item The toroidal magnetic field in the disk is determined by the balance between vertical shear (of the order of $(H/r)^2$) and ambipolar diffusion, and 
 \item Angular momentum transport mechanisms other than magnetic braking (such as turbulent viscosity) are negligible.
\end{enumerate}

We believe that the discovery of this new disk structure is a theoretical breakthrough for star and planet formation theory.
The disk structure is determined only by observable parameters such as the central star mass, mass accretion rate, disk temperature, and cosmic-ray ionization rate, without including difficult-to-determine parameters such as the viscous parameter $\alpha$.
Using the analytical solution, we can study the planet formation process in the realistic disk
and evolution of the magnetic flux during protostellar evolution.
In the future, we will discuss the broad implications of this disk model for the formation and evolution of protostars and planets.

\subsection{Assumptions employed in the dust growth model and their uncertainty}

Our simulations make several simplifications to the dust growth and dust size distribution.
The largest simplification is the representative size approximation for dust growth in which we assume that the representative size corresponds to the peak dust size of the mass distribution (note that the peak of the dust mass distribution corresponds to the maximum dust size if $q<4$).
Our approximated equation for dust growth can be derived from the coagulation equation (for derivation, see \citep{2016A&A...589A..15S}).
In this study, we solve the evolution of the representative size and regard it to be the maximum dust size $a_{\rm max}$, and set the minimum dust size and power as parameters.
Moreover we implicitly assume that the size distribution can be described by a single power law.

More realistically, the time evolution of the dust size distribution should be considered, and the validity of the simplifications employed in this study should be investigated in future more realistic studies.
Detailed modeling of dust fragmentation may be important because the dust fragmentation can cause a variety of dust size distributions \citep{2011A&A...525A..11B}.
In particular, it is possible to have a large number of small dust grains \citep{2018ApJ...869L..45B}. If this is the case,
the adsorption of charged particles by dust grains is not negligible.

However, our claim, "the disk structure converges to the analytical solution once the dust has grown sufficiently  and the adsorption of charged particles by the dust grains becomes negligible", remains valid regardless of the specific details of the dust distribution and dust growth model.
This is because the essential physics required for the disk to converge to the analytical solutions is that  the ambipolar resistivity is  determined by the balance between ionization and recombination and can be written as $\eta_A = B^2/(C \gamma  \rho^{3/2})$.
In this sense, our results are universal.

  \subsection{Comparisons with previous studies}

  Recently, \citet{2023MNRAS.518.3326L} performed spherically symmetric 1D simulations of collapsing cloud core with considering the coagulation and fragmentation of dust grains. They also calculated the change of resistivities due to the dust growth. They pointed out that
dust growth is a critical process for the resistivity in the protostellar evolution. Furthermore, they also pointed out that dust fragmentation if it happens strongly affects the magnetic resistivities profiles. 

\citet{2023A&A...670A..61M} investigated the time evolution of the collapse of the cloud cores until about $1000$ yr after the formation of the first cores with 3D simulations that consider the dust growth.
They found that the grain sizes reach more than 100 $\mum$ in the inner dense region only in $1000$ yr, and the dust growth significantly affects the resistivities. The timescale of dust growth is consistent with our simulations.

In contrast to those previous studies, we investigated the disk evolution for a longer time after protostar formation with 3D simulations.
In particular, the dominant gravitational source in our simulations is the central protostar (sink) and the gas rotation becomes Keplerian, which is necessary for the simulation results to converge to an analytical solution (see Appendix A).  Thus, future studies should include the numerical treatment of the central star that determines the gravity near the center.

\subsection{Importance of future validation}
The impact of numerical resolution or numerical methods on the simulation results were not explored in the paper because we need (additional) enormous computational costs.
Therefore, it is very important to validate our results (especially convergence to the analytical solution) with other numerical method and/or higher numerical resolution in future studies.

Nevertheless, we expect that the convergence of the simluated disk structures to the analytical solutions is robust for the following reasons.
In our simulations that converged to the analytical solution (thick lines in figure \ref{rad_prof_all_models}),
the scale heights of the disks  were resolved with different numerical resolutions (with $\sim 4$ to $10$ smoothing lengths)
due to different densities at the midplane.  Nevertheless, convergence to the power law is observed in all those simulations.
We think this point indirectly reinforces our claim.

\subsection{Maximum dust size lifted up by the outflow}
Figure \ref{dust_size_mass} as well as our previous study \citep{2021ApJ...920L..35T} shows that the dust-to-gas mass ratio increases when the mean dust size in the disk reaches $a_d\gtrsim 100 \mum$.
This is caused by the selective fall of dust grains from the dust-gas mixture lifted up by the outflow.
The simulations show that, once the dust grows to $\sim 100 \mum$, "ash-fall" phenomenon occurs, in which the dust grains and gas are decoupled in the outflow, and only the dust falls back into the disk.
Hence, the minimum dust size for the protostellar ash-fall is $\sim 100 \mum$.
Then, how large is the maximum dust size that can be lifted by the outflow or the maximum dust size of the "falling ash"?
This is particularly important in explaining the recent observations of the presence of grown dust in the envelope \citep{2009ApJ...696..841K, 2019A&A...632A...5G,2019MNRAS.488.4897V}.

The maximum dust size lifted up by the outflow can be estimated from the following considerations.
For the dust grain to be lifted up by the outflow, the dust grains must couple to the gas at the outflow driving point (or root).
Thus, the stopping time of the dust grains should be less than the orbital period at the root (otherwise, outflow driving causes dust grains to remain in the disk and only the gas is ejected).

To estimate the stopping time, we need the density at the outflow root.
As shown in figure \ref{Fout} and by the observations \citep{2004A&A...426..503W}, the mass ejection rate of the outflow in young protostars (e.g., their age is $t\lesssim 10^5$ yr) is in the range $\dot{M}_{\rm out}=10^{-6}-10^{-5} \msunyear$.
Thus, by assuming that the outflow velocity is comparable to the orbital velocity at the radius of the root \citep{1997ApJ...474..362K},
the density at the root of the outflow $\rho_{\rm dp}$ can be estimated as,
\begin{align}
  \rho_{\rm dp}&=\frac{\dot{M}_{\rm out}}{v_{\rm out} \pi r_{\rm disk}^2}  \\
  &=1.9 \times 10^{-15} \nonumber \\
  &\left(\frac{\dot{M}_{\rm out}}{10^{-5} \msunyear}\right)  \left(\frac{M_{*}}{\rm 0.2 \msun}\right)^{-1/2} \left(\frac{r_{\rm disk}}{50 {\rm AU}}\right)^{-3/2} \gcm, \nonumber
\end{align}
where we assume that $v_{\rm out}=\sqrt{G M_*/r}$ is the Keplerian velocity at disk outer edge $r_{\rm disk}$.
Hence, the stopping time at the root is estimated as 
\begin{align}
  t_{\rm stop}&=  \frac{\rho_{\rm mat} a_d}{\rho_{\rm dp} \sqrt{8/\pi}c_s} \\
  &= 3.3 \times 10^2 \left(\frac{\rho_{\rm mat}}{ 2 \g}\right)\left(\frac{a_{d}}{ 5 \mm}\right) \nonumber \\
  &\left(\frac{M_{*}}{\rm 0.2 \msun}\right)^{1/2} \left(\frac{r_{\rm disk}}{50 {\rm AU}}\right)^{12/7} \left(\frac{\dot{M}_{\rm out}}{10^{-5} \msunyear}\right)^{-1}  {\rm yr}, \nonumber
\end{align}
where the sound velocity and temperature is assumed to be $c_s=190 (T/10 {\rm K})^{1/2} \ms$ and $T=150(r/\au)^{-3/7} {\rm K}$, respectively.
Then, the ratio of the stopping time to the orbital period at $r_{\rm disk}$ is calculated as,
\begin{align}
  \frac{t_{\rm stop}}{t_{\rm orb}}&=0.41 \left(\frac{\rho_{\rm mat}}{ 2 \g}\right) \\
  &\left(\frac{a_{d}}{ 5 \mm}\right) \left(\frac{M_{*}}{\rm 0.2 \msun}\right) \nonumber \\
  &\left(\frac{r_{\rm disk}}{50 {\rm AU}}\right)^{3/14} \left(\frac{\dot{M}_{\rm out}}{10^{-5} \msunyear}\right)^{-1}, \nonumber
\end{align}
or $t_{\rm stop} \sim t_{\rm orb}$ is realized when
\begin{align}
  a_{d} &\sim 1.2 \left(\frac{\rho_{\rm mat}}{ 2 \g}\right)^{-1}\left(\frac{M_{*}}{\rm 0.2 \msun}\right)^{-1} \\
  &\left(\frac{r_{\rm disk}}{50 {\rm AU}}\right)^{-3/14} \left(\frac{\dot{M}_{\rm out}}{10^{-5} \msunyear}\right)\cm. \nonumber
\end{align}
This indicates that the dust size of at maximum $a_d\sim 1 \cm$ can be entrained by the outflow with $\dot{M}_{\rm out} \sim  10^{-5} \msunyear$  from the disk with a size of $\sim 50$ AU.
This size is larger than the wavelength of sub-millimeter observations such as with ALMA and may cause the decrease of the spectral index of dust opacity in the outflow and the envelope.
Thus, ``ash-fall" can explain the presence of grown dust in the envelope suggested by the observations.




\section*{Acknowledgments}
We thank Dr. Shinsuke Takasao and Mr. Ryoya Yamamoto for the fruitful discussion.
The computations were performed on the Cray XC50 system at CfCA of NAOJ.
This work is supported by JSPS KAKENHI  grant number 18H05437, 18K13581, 18K03703.


\appendix
\section{Analytic solution of steady state disks with magnetic braking and ambipolar diffusion}
In this section, we derive the power laws of steady-state circumstellar disks which is determined by the angular momentum removal by magnetic braking and magnetic field structure determined by the balance between gas advection and ambipolar diffusion.
We start from MHD equation with ambipolar diffusion,
\begin{eqnarray}
  \frac{\partial \rho}{\partial t}+\nabla \cdot (\rho \vel)&=&0,\\
  \rho \left( \frac{\partial \vel}{\partial t} +\vel \cdot \nabla \vel \right)&=&-\rho \nabla \Phi -\nabla p +\frac{1}{4 \pi} (\nabla \times \magB)\times \magB, \nonumber \\ \\
  \frac{\partial \magB}{\partial t} &=& \nabla \times \left(\vel \times \magB \right) \\
  &-& \nabla \times \left( \frac{\eta_A}{|\magB|^2} ((\nabla \times \magB)\times \magB) \times \magB \right), \nonumber \\
  \nabla \cdot \magB&=&0.
\end{eqnarray}
In this appendix,  $\rho$ denotes the gas density,
$\vel$ denotes the gas velocity,
$p$ denotes the gas pressure,
$\Phi$ denotes the gravitational potential,
$\magB$ denotes the magnetic field.
Hereafter, we assume the steady state i.e., $\frac{\partial}{\partial t}=0$.

Because we focus on the structure of the circumstellar disk,
we make the assumptions below following \citet{2012MNRAS.424.2097G,2013MNRAS.430..822G}.
We introduce a small dimensionless parameter $\epsilon=O(H/r)$, where $H$ denotes the gas scale height of the disk.
When the disk self-gravity is negligible, the gravitational potential is expanded as
\begin{eqnarray}
  \Phi(r,z)=\Phi_0(r)+\frac{1}{2}\Phi_2(r) z^2+O\left(\left(\frac{z}{r}\right)^4\right) \nonumber,\\
  \therefore   \Phi(r,\zeta)=\Phi_0(r)+\epsilon^2 \frac{1}{2}\Phi_2(r)\zeta^2+O(\epsilon^4),
\end{eqnarray}
where we introduces the rescaled vertical coordinate $\zeta=\epsilon^{-1} z$, and
$\Phi_0(r)=-(GM/r)$ and $\Phi_2(r)=(GM/r^3)$.
This dependence of the gravitational potential on $\epsilon$
is the guiding principle that determines the order of other physical quantities.

We assume that the vertical gravity is balanced by thermal pressure at the leading order.
Thus, the scaling of pressure can be assumed to be
\begin{eqnarray}
  p(r,\zeta)=\epsilon^2 p_2(r,\zeta)+\epsilon^4 p_4(r,\zeta)+O(\epsilon^6).
\end{eqnarray}

We assume the scaling of the density to be
\begin{eqnarray}
  \rho(r,\zeta)=\rho_0(r,\zeta)+\epsilon^2 \rho_2(r,\zeta)+O(\epsilon^4).
\end{eqnarray}
Then, the sound velocity $c_s$ scales as
\begin{eqnarray}
  c_s(r)=\epsilon c_{s,1}(r)+\epsilon^3 c_{s,3}(r)+O(\epsilon^5).\\
\end{eqnarray}
Here, we assume the disk to be vertically isothermal.

The leading order of the radial velocity is assumed to be $\epsilon c_s$, and thus,
\begin{eqnarray}
  v_r(r,\zeta)=\epsilon^2 v_{r,2}(r,\zeta)+\epsilon^4 v_{r,4}(r,\zeta)+O(\epsilon^6).
\end{eqnarray}

The leading order of the azimuthal velocity is assume to be balanced with the leading order of $\Phi(r,\zeta)$ and hence,
\begin{eqnarray}
  v_\phi(r,\zeta)=r \Omega_0(r)+ \epsilon^2 v_{\phi,2}(r,\zeta)+\epsilon^4 v_{\phi,4}(r,\zeta)+O(\epsilon^6).
\end{eqnarray}

The leading order of the vertical velocity is assumed to be smaller than $v_r$,
\begin{eqnarray}
  v_z(r,\zeta)=\epsilon^3 v_{z,3}(r,\zeta)+\epsilon^5 v_{r,5}(r,\zeta)+O(\epsilon^7).
\end{eqnarray}

For the magnetic field, we assume that the vertical magnetic field
is the dominant component and the scaling of magnetic field is assumed to be 
\begin{eqnarray}
  B_r(r,\zeta)&=&\epsilon^2 B_{r,2}(r,\zeta)+\epsilon^4 B_{r,4}(r,\zeta)+O(\epsilon^6),\\
  B_\phi(r,\zeta)&=&\epsilon^2 B_{\phi,2}(r,\zeta)+\epsilon^4 B_{\phi,4}(r,\zeta)+O(\epsilon^6),\\
  B_z(r,\zeta)&=&\epsilon B_{z,1}(r)+\epsilon^3 B_{z,3}(r,\zeta)+O(\epsilon^5),
\end{eqnarray}
where the leading term of $B_z(r,\zeta)$ does not depend on $\zeta$ because the leading order of the divergence free condition gives,
\begin{eqnarray}
 \partial_\zeta B_{z,1}=0.
\end{eqnarray}
The underlying assumption that leads to this ordering is that the leading order of Alfven velocity ($\propto B_z/\sqrt{\rho_g}=O(\epsilon)$) is the order of the sound velocity.

  Once the dust grains have grown sufficiently and the adsorption of charged particles by grains becomes negligible, the ionization degree is determined by the balance between cosmic-ray ionization and gas-phase recombination. In this case, $\eta_A$ is given as
\begin{align}
  \eta_A&=\frac{\magB^2}{4 \pi C \gamma \rho^{3/2}} \nonumber \\
  &=\epsilon^2 \eta_{A,2} + O(\epsilon^4).
\end{align}
Here $C$ is given as 
\begin{align}
C=\sqrt{\frac{m_i^2 \zeta_{\rm CR}}{m_g \beta_r}},
\end{align}
where $m_i$ and $m_g$ are the mass of ion and neutral particles and we assume $m_i=29 m_p$ and $m_g=2.34 m_p$ assuming that the major ion is HCO$^+$ where $m_p$ is the proton mass. $\zeta_{\rm CR}$ is the cosmic ray ionization rate. $\beta_r$ is the recombination rate and assumed to be
\begin{align}
 \label{beta_UMIST}
\beta_r=\beta_{r,0} \left(\frac{T}{300 {\rm K}}\right)^{-0.69} \sim \beta_{r, 0} \left(\frac{T}{300 {\rm K}}\right)^{-7/10},
\end{align}
where $\beta_{r,0}=2.4 \times 10^{-7} {\bf cm^3 s^{-1}}$ taken from UMIST database \citep{2013A&A...550A..36M}.
\begin{align}
\gamma=\frac{\langle \sigma v \rangle_{in}}{(m_g+m_i)},
\end{align}
where $\langle \sigma v \rangle_{in}$ is the rate coefficient for collisional momentum transfer between ions and neutrals.
We assume  $\langle \sigma v \rangle_{in} = 1.3 \times 10^{-9} {\rm cm^3 s^{-1}}$ which is calculated from the Langevin rate \citep{2008A&A...484...17P}.
Hence $\eta_A$ has the weak temperature dependence of approximately $\propto T^{-\frac{7}{20}}$.

The leading order of $\eta_A$ is $\epsilon^2$ because of the leading order of $B_z(r,\zeta)$ and $\rho(r,\zeta)$ are  $O(\epsilon)$ and  $O(1)$, respectively.

The radial components of the equation of motion at the leading order,
\begin{eqnarray}
  -\rho_0 r \Omega_0^2=-\rho_0 \partial_r \Phi_0,
\end{eqnarray}
gives
\begin{eqnarray}
  \Omega_0=\sqrt{\frac{G M}{r^3}}.
\end{eqnarray}

The vertical components of equation of motion at the leading order
\begin{eqnarray}
\rho_0 \Phi_2 \zeta =-\partial_\zeta p_2,
\end{eqnarray}
leads
\begin{eqnarray}
\rho_0=\frac{\Sigma_0}{\sqrt{2 \pi} H_1}\exp\left(-\frac{\zeta^2}{2H_1^2}\right)\equiv \tilde{\rho}\exp\left(-\frac{\zeta^2}{2H_1^2}\right),
\end{eqnarray}
where $H_1(r)=c_{s,1}/\Phi_2^{1/2}=c_{s,1}/\Omega_0$, where $c_{s,1}$ is the vertically
isothermal sound velocity defined by $p_2=c_{s,1}^2\rho_0$ and $H_1(r)$ is related to the scale height as $H= \epsilon H_1+O(\epsilon^3)$.
$\Sigma_0$ is given as $\Sigma_0=\int_{-\infty}^\infty \rho_0 d\zeta$.

The second order of the radial and azimuthal components 
of the equation of motion and of the induction equation are written as
\begin{eqnarray}
  \label{eqs_second_order_first}
  -2\rho_0 \Omega_0 v_{\phi,2}&=&-\frac{1}{2} \rho_0 \partial_r\Phi_2 \zeta^2 \nonumber \\
  &-&\partial_r\left(p_2+\frac{B_{z,1}^2}{8\pi}\right)+\frac{B_{z,1}}{4 \pi}\partial_\zeta B_{r,2}, \\
  \rho_0 v_{r,2} \frac{1}{r} \partial_r(r^2 \Omega_0) &=&\frac{B_{z,1}}{4 \pi}\partial_\zeta B_{\phi,2},\\
  0=B_{z,1}\partial_\zeta v_{r,2} &+&\partial_\zeta[\eta_{A,2} (\partial_\zeta B_{r,2}-(\partial_r B_{z,1})]. \\
  \label{eqs_second_order_last}
  0=B_{r,2} r \partial_r \Omega_0  &+& B_{z,1} \partial_\zeta v_{\phi,2}+\partial_\zeta (\eta_{A,2} \partial_\zeta B_{\phi,2})
\end{eqnarray}
These equations correspond to equation (28) to (31) of
\citet{2012MNRAS.424.2097G}, if we assume viscous parameter $\alpha$ to be $0$.

Then, we rescale the vertical coordinate with
\begin{eqnarray}
\hat{z} \equiv \frac{\zeta}{H_1}=\frac{z}{H}.
\end{eqnarray}

To evaluate the single power law for each physical quantity,
we need to further simplify the equations (\ref{eqs_second_order_first}) to (\ref{eqs_second_order_last}).
Here, we assume that the radial thermal pressure gradient is much larger
than the magnetic pressure gradient, and the terms with $B_{r,2}$ can be neglected.

Then equations (\ref{eqs_second_order_first}) to (\ref{eqs_second_order_last}) are rewritten as
\begin{eqnarray}
  \label{eqs_second_order_simplified_first}
  -2\rho_0 \Omega_0 v_{\phi,2}&=&\frac{3}{2 r} \rho_0 \Omega_0^2 H_1^2 \hat{z}^2-\partial_r(\rho_0 c_{s,1}^2),\\
  \frac{1}{2} \rho_0 v_{r,2} \Omega_0 &=&\frac{B_{z,1}}{4 \pi H_1}\partial_{\hat{z}} B_{\phi,2},\\
  B_{z,1}\partial_{\hat{z}} v_{r,2} &=& (\partial_{\hat{z}} \eta_{A,2}) (\partial_r B_{z,1}), \\
  B_{z,1} \partial_{\hat{z}} v_{\phi,2}&=&\frac{1}{H_1} \partial_{\hat{z}} (\eta_{A,2} \partial_{\hat{z}} B_{\phi,2}).
  \label{eqs_second_order_simplified_last}
\end{eqnarray}

Furthermore, from the conservation of the mass, we have
\begin{eqnarray}
  \label{eqs_mass_accretion}
-2\pi r  \int^{H}_{-H} \rho v_r d z  = \dot{M} \equiv \epsilon^3 \dot{M}_{3} +O(\epsilon^5).
\end{eqnarray}
where we approximated the integral range from $-H$ to $H$ instead of from $-\infty$ to $\infty$
because of the vertical expansion below.
$\dot{M}$ is the mass accretion rate within the disk which is assumed to be constant and
$\dot{M}_{3}$ is its leading term.

Using $\hat{z}$, we consider the vertical expansion.
Because the azimuthal component of the magnetic field should be odd with respect to the midplane,
we assume a vertical dependence of the magnetic field as
\begin{eqnarray}
  B_{\phi,2}=B_{\phi,\hat{z}1} \hat{z} +B_{\phi,\hat{z}3} \hat{z}^3  +O(\hat{z}^5).
\end{eqnarray}
Note that $B_{r,2}$ has disappeared from the equations and cannot be determined from our assumptions above.

The radial and azimuthal components of the velocity should be even with respect to the midplane.
Thus, we assume the vertical dependence of the velocity as
\begin{eqnarray}
  v_{r,2}=v_{r,\hat{z}0} +v_{r,\hat{z}2} \hat{z}^2 + O(\hat{z}^4),\\
  v_{\phi,2}=v_{\phi,\hat{z}0} +v_{\phi,\hat{z}2} \hat{z}^2  +O(\hat{z}^4).
\end{eqnarray}

Then we assume the power law for the vertical magnetic field $B_z$ and midplane density $\rho_{\rm mid}\equiv \rho(r,z=0)$
with respect to the radius as 
\begin{eqnarray}
  B_z(r)=B_{\rm z, ref}\left(\frac{r}{r_{\rm ref}}\right)^{D_{B_z}},\\
  \rho_{\rm mid}(r)=\rho_{\rm mid, ref}\left(\frac{r}{r_{\rm ref}}\right)^{D_{\rho_g}}.
\end{eqnarray}

To be consistent with our simulations, we assume the polytropic relation for the sound velocity
\begin{eqnarray}
  \label{cs_polytropic}
  c_s(r)=c_{\rm s,ref} \left(\frac{\rho_{\rm mid}(r)}{\rho_c}\right)^{1/3},
\end{eqnarray}
     and temperature
\begin{eqnarray}
\label{Tgas_polytropic}
  T(r)=T_{\rm ref} \left(\frac{\rho_{\rm mid}(r)}{\rho_c}\right)^{2/3},
\end{eqnarray}
where we assume $T_{\rm ref}=10$ K. Note that although we assume the temperature distribution in this Appendix, 
the model described here can be applied to different temperature distribution.

From the analytical form, $\eta_A$ is vertically expanded as 
\begin{eqnarray}
  \eta_{A,2}&=&\frac{B_{z,1}(r)^2}{C \gamma \rho_0(r,\hat{z})^{3/2}} =B_{z,1}^2 \tilde{\rho}^{-(3/2+7/30)} \exp((\frac{3}{4}+\frac{7}{60})\hat{z}^2) \nonumber \\
  &=& (C \gamma)^{-1} B_{z,1}^2 \tilde{\rho}^{-3/2}\left(1+(\frac{3}{4}+\frac{7}{60})\hat{z}^2+O(\hat{z}^4)\right),
\end{eqnarray}
where the factor $\frac{7}{60}$ comes from the temperature dependence of the recombination rate.

By substituting these power laws and taking the leading terms with respect to the $\epsilon$ of equation (\ref{eqs_mass_accretion}),
we obtain the solutions of the equations (\ref{eqs_second_order_simplified_first}) to (\ref{eqs_mass_accretion}),

\begin{eqnarray}
  v_{\phi,\hat{z}2}&=&\frac{82}{150} v_{\phi,\hat{z}0},\\
  v_{r,\hat{z}2}&=&\frac{1}{2} v_{r,\hat{z}0},\\
  B_{\phi,\hat{z}3}&=&0,
\end{eqnarray}
and, 
\begin{align}
 \label{solution_first}
 \rho_{\rm mid}(r)&=5.1 \times 10^{-1}\dot{M}^{\frac{15}{41}} \rho_c^{\frac{37}{82}} \Omega_0^{\frac{45}{41}} m_g^{\frac{15}{82}} m_i^{-\frac{15}{41}}\nonumber \\
 &\zeta_{\rm CR}^{-\frac{15}{82}} \beta_{r,0}^{-\frac{35}{164}} \gamma^{-\frac{15}{41}} c_{\rm s,ref}^{-\frac{45}{41}}  \left(\frac{r}{r_{\rm ref}}\right)^{-\frac{135}{82}},\\
   \label{solution_second}
  B_z(r)&= 3.6 \times 10^{-1} \dot{M}^{\frac{47}{82}} \rho_c^{\frac{23}{164}} \Omega_0^{\frac{59}{82}} m_g^{-\frac{35}{164}} m_i^{\frac{35}{82}} \nonumber \\
  &\zeta_{\rm CR}^{-\frac{35}{164}}  \beta_{r,0}^{-\frac{35}{164}} \gamma^{\frac{35}{82}} c_{\rm s,ref}^{-\frac{59}{82}} \left(\frac{r}{r_{\rm ref}}\right)^{-\frac{177}{164}},\\
  v_\phi(r,z=0)-v_K&=-8.8 \times 10^{-1} \dot{M}^{\frac{10}{41}} \rho_c^{-\frac{15}{41}} \Omega_0^{\frac{30}{41}} m_g^{\frac{5}{41}} m_i^{-\frac{10}{41}} \nonumber \\
  &\zeta_{\rm CR}^{-\frac{5}{41}}  \beta_{r,0}^{\frac{5}{41}} \gamma^{-\frac{10}{41}} c_{\rm s,ref}^{\frac{11}{41}} \left(\frac{H_{\rm ref}}{r_{\rm ref}}\right) \left(\frac{r}{r_{\rm ref}}\right)^{-\frac{49}{82}},\\
  v_{r}(r,z=0)&=- 2.1 \times 10^{-1} \dot{M}^{\frac{21}{41}} \rho_c^{-\frac{11}{41}} \Omega_0^{\frac{22}{41}} m_g^{-\frac{10}{41}} m_i^{-\frac{20}{41}} \nonumber \\
  &\zeta_{\rm CR}^{\frac{10}{41}}  \beta_{r,0}^{-\frac{10}{41}} \gamma^{\frac{20}{41}} c_{\rm s,ref}^{-\frac{22}{41}} \left(\frac{H_{\rm ref}}{r_{\rm ref}}\right) \left(\frac{r}{r_{\rm ref}}\right)^{-\frac{25}{82}},\\
  \label{solution_last}
  B_{\phi}(r,z)&= -1.5 \times 10^{-1} \dot{M}^{\frac{35}{82}} \rho_c^{-\frac{23}{164}} \Omega_0^{\frac{105}{82}} m_g^{\frac{35}{164}} m_i^{-\frac{35}{82}} \nonumber \\
  &\zeta_{\rm CR}^{-\frac{35}{164}}  \beta_{r,0}^{\frac{35}{164}} \gamma^{-\frac{35}{82}} c_{\rm s,ref}^{-\frac{23}{82}} \left(\frac{H_{\rm ref}}{r_{\rm ref}}\right) \left(\frac{z}{H}\right) \left(\frac{r}{r_{\rm ref}}\right)^{-\frac{233}{164}},
\end{align}
where $H_{\rm ref}\equiv c_{\rm s, ref}/\Omega(r_{\rm ref})$.

The following estimates are obtained by substituting numerical values for the parameters.
\begin{align}
  \rho_{\rm mid}(r)&=1.7\times 10^{-12} \nonumber \\
  &\left(\frac{\dot{M}}{10^{-5} \msun {\rm yr}^{-1}}\right)^{\frac{15}{41}} \left(\frac{\rho_c}{10^{-13} \gcm}\right)^{\frac{37}{82}} \nonumber \\
  &\left(\frac{M_{\rm star}}{0.2 \msun}\right)^{\frac{45}{82}} \left(\frac{\zeta_{\rm CR}}{10^{-17} s^{-1}}\right)^{-\frac{15}{82}} \nonumber \\
  &\left(\frac{c_{\rm s,ref}}{190 \ms}\right)^{-\frac{45}{41}} \left(\frac{r}{10 {\rm AU}}\right)^{-\frac{135}{82}} \gcm,\\
  B_z(r)&= 2.5\times 10^{-1} \left(\frac{\dot{M}}{10^{-5} \msun {\rm yr}^{-1}}\right)^{\frac{47}{82}} \nonumber\\
  &\left(\frac{\rho_c}{10^{-13} \gcm}\right)^{\frac{23}{164}} \left(\frac{M_{\rm star}}{0.2 \msun}\right)^{\frac{59}{164}} \left(\frac{\zeta_{\rm CR}}{10^{-17} s^{-1}}\right)^{\frac{35}{164}} \nonumber \\
  &\left(\frac{c_{\rm s,ref}}{190 \ms}\right)^{-\frac{59}{82}} \left(\frac{r}{10 {\rm AU}}\right)^{-\frac{177}{164}} {\rm G},\\
  v_\phi (r,z)  &= v_K -80 \left(1+\frac{82}{150} \left(\frac{z}{H}\right)^2\right)\left(\frac{\dot{M}}{10^{-5} \msun {\rm yr}^{-1}}\right)^{\frac{10}{41}} \nonumber \\
  &\left(\frac{\rho_c}{10^{-13} \gcm}\right)^{-\frac{15}{41}} \left(\frac{M_{\rm star}}{0.2 \msun}\right)^{\frac{15}{41}} \left(\frac{\zeta_{\rm CR}}{10^{-17} s^{-1}}\right)^{-\frac{5}{41}} \nonumber \\
  &\left(\frac{c_{\rm s,ref}}{190 \ms}\right)^{\frac{11}{41}} \left(\frac{r}{10 {\rm AU}}\right)^{-\frac{49}{82}} \ms,\\
  v_{r}(r,z)&= -110 \left(1+\frac{1}{2} \left(\frac{z}{H}\right)^2\right)\left(\frac{\dot{M}}{10^{-5} \msun {\rm yr}^{-1}}\right)^{\frac{21}{42}} \nonumber \\
  &\left(\frac{\rho_c}{10^{-13} \gcm}\right)^{-\frac{11}{41}} \left(\frac{M_{\rm star}}{0.2 \msun}\right)^{\frac{11}{41}} \left(\frac{\zeta_{\rm CR}}{10^{-17} s^{-1}}\right)^{\frac{22}{41}} \nonumber \\
  &\left(\frac{c_{\rm s,ref}}{190 \ms}\right)^{-\frac{22}{41}}  \left(\frac{r}{10 {\rm AU}}\right)^{-\frac{25}{82}} \ms,\\
  B_{\phi}(r,z)&= -2.4\times 10^{-2}  \left(\frac{z}{H}\right) \left(\frac{\dot{M}}{10^{-5} \msun {\rm yr}^{-1}}\right)^{\frac{35}{82}} \nonumber \\
  &\left(\frac{\rho_c}{10^{-13} \gcm}\right)^{-\frac{23}{164}} \left(\frac{M_{\rm star}}{0.2 \msun}\right)^{\frac{105}{164}} \left(\frac{\zeta_{\rm CR}}{10^{-17} s^{-1}}\right)^{-\frac{35}{164}} \nonumber \\
  &\left(\frac{c_{\rm s,ref}}{190 \ms}\right)^{-\frac{23}{82}} \left(\frac{r}{10 {\rm AU}}\right)^{-\frac{233}{164}} {\rm G}.
\end{align}
These equations reproduce our simulation results very well.

So far, we have derived the solution from the basic equations with the assumptions explicitly stated,
which, however, may be intuitively difficult to understand.
The aforementioned power laws can be derived from a simple and intuitive extension of the viscous accretion disk model.
Besides the numerical factors, the analytic solutions can also be derived by the following equations:
\begin{eqnarray}
  \label{a7}
  -2 \pi r v_r \Sigma =\dot{M},\\
  \label{a8}
  v_\phi =\sqrt{\frac{G M_c}{r}}\equiv r \Omega,\\
  \label{a9}
  H=\frac{c_s}{\Omega},\\
  \label{a10}
   v_r =-\frac{B_z B_{\phi, s}}{\pi \Sigma \Omega},\\
  \label{a11}
  B_z v_r=-\frac{\eta_A}{r} B_z,\\
  \label{a12}
  B_{\phi, s} =\left( \frac{H}{r}\right)^2 \frac{B_z H}{\eta_A} v_\phi.
\end{eqnarray}
The equations (\ref{a7}) to (\ref{a10}) have a similar form
of the standard viscous accretion disk model \citep{1973A&A....24..337S,1974MNRAS.168..603L},
in which equation (\ref{a10}) is
\begin{eqnarray}
v_r= -\frac{3}{2} \frac{\alpha c_s H}{r},
\end{eqnarray}
where $\alpha$ is the Shakura-Sunyaev viscous parameter.

In our model, equations  (\ref{a11}) and  (\ref{a12}) describe the radial and azimuthal balance
between the magnetic field advection by the gas motion and magnetic field drift by ambipolar diffusion.
Note that the right hand side of equation (\ref{a12}) has the factor of $(H/r)^2$, reflecting that the balance between the vertical shear motion (not  Keplerian rotation itself) and magnetic field drift due to ambipolar diffusion determines the toroidal magnetic field $B_{\phi}$.
This is the key to deriving the disk structure we have identified.

Instead of equation (\ref{a12}), it is often assumed that
\begin{eqnarray}
  \label{a12_prev}
  B_{\phi, s} = \frac{B_z H}{\eta_A} v_\phi,
\end{eqnarray}
which corresponds to the assumption that the Keplerian rotation balances the magnetic field drift by ambipolar diffusion \citep{2002ApJ...580..987K,2012MNRAS.422..261B,2016ApJ...830L...8H}.
We found that this relation leads to considerably large $v_r$ (of the order of the Keplerian velocity), which is inconsistent with the simulation results.
The reason why this relation is inappropriate for the circumstellar disk is that
at the leading order of $\epsilon$,  the gravity does not depend on $z$ and is canceled out by the centrifugal force in the circumstellar disk.
Thus, we should consider the balance between rotation and field drift at the order of $\epsilon^2$ to estimate the toroidal magnetic field in the circumstellar disk (see the derivations above for details  and see also equation (27) of \citet{2021MNRAS.508.2142X}). 

The solutions given by equations (\ref{solution_first}) to (\ref{solution_last}) well reproduce our three dimensional simulations.
Furthermore, they are specified only by the central star mass, mass accretion rate, equation of state (or gas temperature), and ionization and recombination rate,
and {\it do not} contain the viscous $\alpha$ parameter, which is usually extremely difficult to determine.
Therefore, we believe that the analytical solutions are useful for investigating the longer-term evolution of circumstellar disks than we have investigated in this paper by the simulations.

\section{dust size dependence of ambipolar resistivity}
\begin{figure*}
    \includegraphics[clip,trim=0mm 0mm 0mm 0mm,width=100mm,angle=-90]{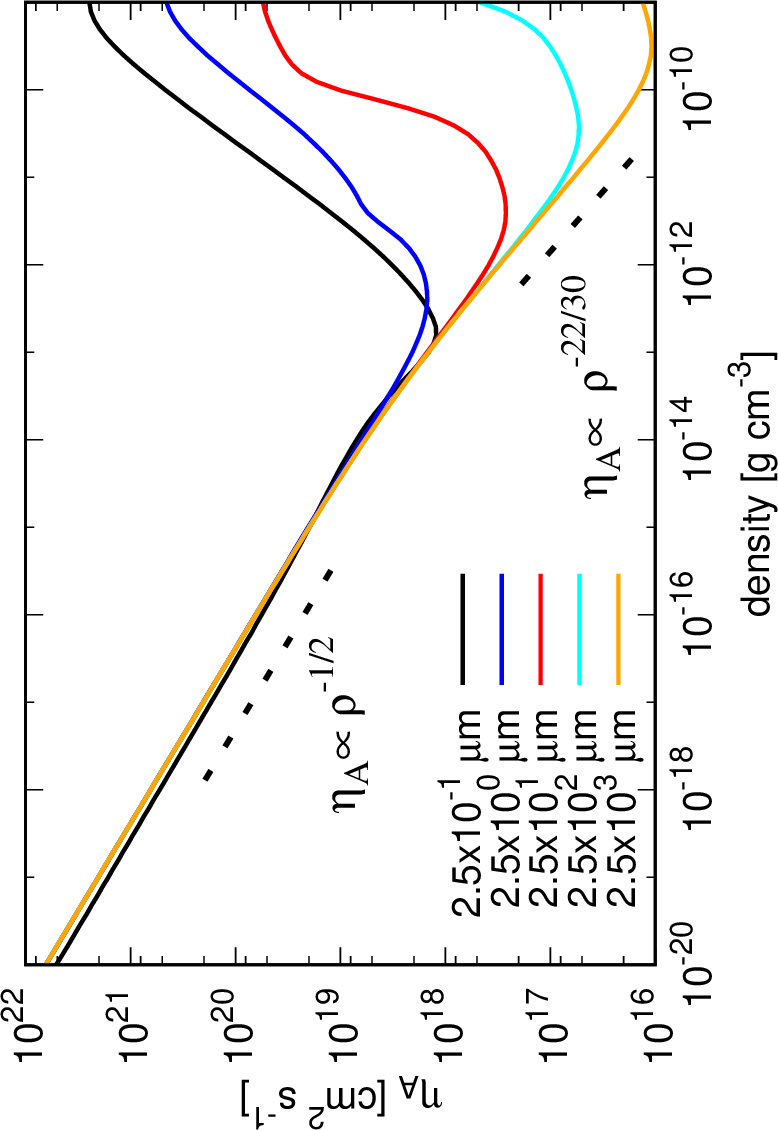}
    \caption{
     $\eta_{\rm A}$ as a function of the density with parameters of ModelA100Q25 in which we assume $a_{\rm min}=100 \nm$, $q=2.5$, and $\zeta_{\rm CR}=10^{-17} {\rm s^{-1}}$
      The black, blue, red, cyan, orange lines show the results of $a_{\rm max}=2.5\times 10^{-1} \mum,2.5\times 10^{0} \mum,2.5\times 10^{1} \mum, 2.5\times 10^{2} \mum, 2.5\times 10^{3} \mum$, respectively.
}
\label{etaA_amax}
\end{figure*}

It would be insightful to see how  $\eta_A$ changes depeding on the maximum dust size with a simple one zone model.
Figure \ref{etaA_amax} shows the $\eta_A$ dependence on the maximum dust size with the parameter adopted in ModelA100Q25 (the fiducial model).

Here, we assume that the temperature
is given as $T=10(1+\gamma (\rho_{\rm g}/\rho_c)^{(\gamma-1)}) ~{\rm K}$, where $\gamma=5/3$ and $\rho_c=4\times 10^{-13} \gcm$.
The magnetic field is given as $0.2 n_{\rm g}^{1/2} \mu G$  \citep[i.e., assuming flux freezing; see e.g.,][]{2002ApJ...573..199N}.  $\zeta_{\rm CR}=10^{-17} {\rm s^{-1}}$.
With these assumptions of the temperature and magnetic field, and once the adsorption of charged particles by grains becomes negligible, $\eta_A$ obeys
\begin{eqnarray}
\eta_A =\frac{B^2}{4 \pi C \gamma \rho^{\frac{3}{2}}} &\propto&
  \begin{cases}
    \rho^{-\frac{1}{2}} & (\rho<\rho_c)\\
    \rho^{-\frac{22}{30}} & (\rho>\rho_c),
  \end{cases}
\end{eqnarray}
where the temperature dependence of $C$ is included (equation (\ref{beta_UMIST})).

We can see that $\eta_A$ is well described by this power law in $\rho<10^{-11} \gcm$ (i.e., the simulated disk density region)
once $a_{\rm max} \gtrsim 250 \mum$.
Therefore, our assumption in Appendix A seems justified when the dust size exceeds $\gtrsim 100 \mum$ (at least for ModelA100Q25).
See \citet{2022ApJ...934...88T} for the results with a wider variety of parameters.

\bibliography{article}

\end{document}